\documentclass[preprint,12pt]{elsarticle}

\usepackage[utf8]{inputenc}
\usepackage[margin=1in]{geometry}
\usepackage{setspace}
\usepackage{tabularx}
\usepackage{array}
\usepackage{booktabs}
\usepackage{multirow}
\usepackage{longtable}
\usepackage{colortbl}
\usepackage{hhline}
\usepackage{nicematrix}
\usepackage{titlesec}
\usepackage{makecell}
 \usepackage{float} 
\usepackage{amssymb}
\usepackage{booktabs}
\usepackage{siunitx}
\usepackage{subcaption}
\usepackage{placeins}
\usepackage{hyperref}
\usepackage{graphicx}
\usepackage{xcolor}
\definecolor{SeaGreen}{rgb}{0.18,0.55,0.34}
\definecolor{PineGreen}{rgb}{0.20,0.88,0.24}
\usepackage{subcaption}
\usepackage{subcaption}
\usepackage{amsmath}
\usepackage{xurl}

\usepackage{booktabs}
\usepackage[table]{xcolor}
\usepackage{array}
\usepackage{xr}
\usepackage[ruled]{algorithm2e}
\DontPrintSemicolon
\SetAlgoNoLine
\usepackage{titlesec}
\titleformat{\paragraph}
  {\normalfont\normalsize\bfseries}{\theparagraph}{1em}{}
\titlespacing*{\paragraph}
  {0pt}{3.25ex plus 1ex minus .2ex}{1.5ex plus .2ex}

\usepackage[numbers]{natbib} 
\usepackage{xurl}
\usepackage{hyperref}
\hypersetup{
  colorlinks   = true,
  urlcolor     = blue,
  linkcolor    = blue,
  citecolor    = red
}
\usepackage[dvipsnames]{xcolor}
\usepackage[normalem]{ulem}
\usepackage{setspace}
\definecolor{DarkReviewerGreen}{RGB}{0,110,70}

\newcommand{\revone}[3]{#3}
\newcommand{\revtwo}[3]{#3}
\usepackage{lineno}
\usepackage{blindtext}
\SetAlCapFnt{\small\bfseries} % Font for the Label
\SetAlCapNameFnt{\small}      % Font for the Title text
\journal{Journal of Biomedical Informatics}

\begin{document}
\onehalfspacing
\begin{frontmatter}

\title{Typical Healthcare Pathways as a Basis for Admixture Modeling of Patient Trajectories}

% Authors
\author[inst1,inst2]{Maryam Farhadizadeh\corref{cor1}}
            
\ead{maryam.farhadizadeh@uniklinik-freiburg.de}
\author[inst2,inst3]{Carola S. Heinzel}
\author[inst4]{August Sigle}
\author[inst2,inst5]{Harald Binder}
\author[inst6]{Frederik Wenz}
\author[inst7,inst8]{Jan Hasenauer}
\author[inst2,inst3]{Peter Pfaffelhuber}
\author[inst1,inst2]{Nadine Binder}
\cortext[cor1]{Corresponding author}

\affiliation[inst1]{
  organization={Institute of General Practice/Family Medicine, Medical Center and Faculty of Medicine, University of Freiburg},
  addressline={Elsässer Str. 4d},
  postcode={79110},
  city={Freiburg},
  country={Germany}
}
\affiliation[inst2]{
  organization={Freiburg Center for Data Analysis, Modeling and AI, University of Freiburg},
  addressline={Ernst-Zermelo-Str. 1},
  postcode={79104},
  city={Freiburg},
  country={Germany}
}
\affiliation[inst3]{
  organization={Mathematical Institute, Division of Mathematical Stochastics, University of Freiburg},
  addressline={Ernst-Zermelo-Str. 1},
  postcode={79104},
  city={Freiburg},
  country={Germany}
}
\affiliation[inst4]{
  organization={Department of Urology, Faculty of Medicine and Medical Center, University of Freiburg},
  addressline={Hugstetter Str. 55},
  postcode={79106},
  city={Freiburg},
  country={Germany}
}        
\affiliation[inst5]{
  organization={Institute of Medical Biometry and Statistics, Faculty of Medicine and Medical Center, University of Freiburg},
  addressline={Stefan-Meier-Straße 26},
  postcode={79104},
  city={Freiburg},
  country={Germany}
}        
\affiliation[inst6]{
  organization={Faculty of Medicine and Medical Center, University of Freiburg},
  addressline={Breisacher Str. 153},
  postcode={79110},
  city={Freiburg},
  country={Germany}
}        
\affiliation[inst7]{
  organization={Bonn Center for Mathematical Life Sciences, University of Bonn},
  addressline={Endenicher Allee 64},
  postcode={53115},
  city={Bonn},
  country={Germany}
}        
\affiliation[inst8]{
  organization={Life \& Medical Sciences Institute, University of Bonn},
  addressline={Carl-Troll-Str. 31},
  postcode={53115},
  city={Bonn},
  country={Germany}
}

\begin{abstract}
Background: Understanding whether patients follow similar or distinct patterns of care is important for characterizing clinical practice, identifying patient subgroups and supporting quality improvement. However, routine healthcare trajectories are difficult to compare directly because patients may differ in their diagnostic workup, treatment sequencing, timing of clinical events, and documentation practices. Despite this patient-specific variation, trajectories often contain recurring structural patterns at the cohort level. \\
Methods: To address this challenge, we present a framework that explicitly separates cohort-level typical pathway identification from patient-level inference. At the cohort level, we derive a simplified and interpretable representation of care processes using a rule-based algorithm to identify typical healthcare pathways, resulting in a compact pathway graph. These pathways are then modeled as Markov chains and used as structured components in an admixture model, allowing each patient to be represented as a probabilistic mixture of typical pathways rather than being assigned to a single pathway component. The resulting admixture weights provide a compact, low-dimensional representation of patient trajectories that can be used for downstream analyses such as subgroup characterization. We further assess the stability of the identified pathways and inferred admixture representations across multiple train–test splits.\\
Results: Across train-test splits, the framework demonstrated consistent pathway structures and patient-level mixture patterns. Applied to routine care data from prostate cancer patients undergoing radical prostatectomy, the framework identified clinically plausible and interpretable care patterns and supported the identification of patient subgroups with similar patterns of coded clinical events. \\
Conclusions: Overall, the proposed framework provides an interpretable and stable approach for summarizing treatment pathways and characterizing patient subgroups in real-world practice.
\end{abstract}


\end{frontmatter}

% ==================================================

\section{Introduction}
Understanding how patients move through real-world care is central to evaluating clinical practice, identifying variation, and improving healthcare delivery. Healthcare pathways are often conceptualized as standardized sequences of diagnostic, therapeutic, and follow-up steps for specific conditions, intended to reduce unwarranted variation and support clinical decision-making \cite{aspland2021clinical,lawal2016clinical}. In routine care, however, observed patient trajectories, i.e., the observed sequences of coded clinical events or states for individual patients, frequently deviate from these idealized models \cite{binder2022,xiao2018opportunities}. Patients with the same diagnosis or treatment indication may differ in comorbidities, disease severity, referral history, provider practices, and treatment decisions, resulting in longitudinal, event-based trajectories with branching, recurrence, concurrent events, and varying levels of documentation detail. These characteristics make it challenging to derive compact and interpretable cohort-level representations of care processes from electronic health records (EHRs) \citep{rosa2022modelling}.

As highlighted in methodological overviews of pathway mining and modeling, e.g., \citep{aspland2021clinical}, the analysis of patient trajectories requires the balancing of three key objectives: reducing high-dimensional event logs to manageable representations, preserving temporal and structural progression, and maintaining interpretability for clinical insight.  
This is complicated by the fact that routine care datasets are inherently fragmented. Even within large cohorts, patients frequently follow diverse and only partly overlapping care patterns, giving rise to numerous seemingly distinct trajectories despite a limited number of underlying treatment options \citep{binder2022}. This structural heterogeneity complicates identifying representative care pathways and summarizing pathway dynamics.
Existing methodological approaches address this challenge from different perspectives, including structural reconstruction methods such as process mining and sequence mining, probabilistic abstraction approaches such as topic modeling, clustering-based models of trajectory heterogeneity, including mixtures of Hidden Markov Models, and predictive machine learning methods for EHRs.

Process mining was among the first methodological approaches used to reconstruct care processes from clinical event logs \citep{vanderAalst2011}. Within this framework, workflow discovery methods infer directed process models from observed event sequences, representing clinical activities, transitions, branching structures, and concurrent events in patient care \citep{rosa2022modelling,rismanchian2023data,quintano2019sepsis}. These models provide detailed descriptions of real-world care delivery and can highlight deviations from expected workflows. However, in fine-grained clinical data, they often retain substantial patient-level variability, which can make them difficult to use as compact and interpretable cohort-level summaries. Frequent sequence mining offers a complementary strategy by extracting recurrent subsequences from patient trajectories \citep{perer2015mining}. Although these subsequences are useful for identifying common local patterns, they are typically identified independently and therefore do not necessarily provide a unified model of overall care progression. Thus, these approaches reveal important recurring structure in clinical event data, but only partially address the challenge of reducing heterogeneous trajectories into compact, interpretable pathway representations. Later approaches introduced abstraction at the level of clinical events. Topic modeling techniques adapt methods originally developed for text analysis to healthcare trajectory data. In this setting, clinical events, such as diagnoses, procedures, medications, or laboratory-related events, are treated analogously to words, while care episodes or patient records are treated as documents. Latent Dirichlet Allocation then represents each care episode as a mixture of latent topics, where each topic corresponds to a probability distribution over clinical activities \citep{blei2003latent}. Applied to healthcare data, these topics can be interpreted as recurrent treatment patterns, care themes, or activity profiles that summarize frequently co-occurring events \citep{huang2018probabilistic,huang2019medication,chiudinelli2020mining}. By replacing sparse, high-dimensional event representations with a smaller number of latent activity patterns, topic modeling reduces dimensionality while preserving clinically meaningful semantic structure. Refinements incorporate daily aggregation and process pruning to derive more concise pathway representations \citep{xu2016tcpm}, and temporal extensions associate latent treatment patterns with time distributions to capture longitudinal regularities \citep{li2024temporal}. Although these probabilistic abstractions condense fine-grained event logs into interpretable latent structures, they primarily operate at the level of events or days and often treat structural progression across complete trajectories implicitly. Alternative approaches address heterogeneity through clustering and event aggregation. Clinical events can be grouped using diagnostic taxonomies before patient sequences are modeled via mixtures of Hidden Markov Models \citep{rabiner1989tutorial,najjar2018two}, or trajectories can be partitioned using K-means clustering \citep{macqueen1967some} with edit-distance similarity \citep{levenshtein1966binary}, with cluster-specific templates derived through sequence alignment \citep{funkner_towards_2017}. Although such approaches reduce variability by grouping similar pathways, structural progression patterns remain distributed across clusters or embedded within fitted models rather than represented through a unified structural scaffold.
In parallel, machine learning methods for EHRs have increasingly been developed to predict longitudinal clinical events, disease onset, and outcome risk, often using statistical learning and deep neural architectures designed for sequential data e.g., \citep{muyama2024machine,xiao2018opportunities,lin2021personalized,ye2020predicting,muyama2024drl}. Although these methods can handle high-dimensional and irregular EHR data, their primary objective is usually prediction or latent representation learning rather than explicit construction of interpretable models of care progression.

Across these approaches, a common limitation emerges: Structural reconstruction methods capture care flows but struggle with complexity, abstraction-based and probabilistic models reduce dimensionality but often treat progression implicitly, and clustering-based approaches partition heterogeneity without providing a unified probabilistic representation of trajectories within an interpretable structural framework. The explicit integration of cohort-level pathway structure with probabilistic modeling of individual trajectories therefore remains insufficiently developed.

In this work, we introduce a three-step framework that addresses this limitation by separating cohort-level pathway identification from patient-level inference. First, we identify typical healthcare pathways, i.e., cohort-level progression patterns identified from multiple patient trajectories, using a rule-based algorithm that compresses common sub-paths across patient trajectories into a compact and interpretable pathway graph. This graph provides a descriptive, resolution-controlled representation of shared care structures, capturing dominant progression patterns at a chosen level of granularity without assuming a probabilistic generative model. Second, we use this cohort-level representation for patient-level inference through an admixture model in which the identified typical pathways are represented as pathway-specific Markov chains \citep{norris1998markov}. Drawing on mixed-membership models originally developed in population genetics \citep{pritchard2000,alexander2009fast}, patients are represented as probabilistic mixtures over these fixed cohort-level components rather than being assigned to a single prototypical trajectory. Third, the resulting admixture weights provide a compact, low-dimensional representation of patient trajectories on a common scale, enabling comparison across patients and supporting downstream subgroup exploration. We illustrate the framework using routine care data from prostate cancer patients undergoing radical prostatectomy at the University Medical Center Freiburg.

% \begin{table}[h]
% \caption*{{\bf Statement of significance}}
% \renewcommand{\arraystretch}{1.4}
% \begin{tabularx}{\textwidth}{p{0.38\textwidth} X}
% \hline

% Problem or Issue &
% How can heterogeneous routine healthcare trajectories be summarized in an interpretable way while preserving patient-level variability for subgroup analysis? \\

% What is already known &
% Healthcare pathways derived from routine care data are often highly heterogeneous, making direct comparison between patients difficult. Existing pathway analysis approaches typically focus either on cohort-level summaries or on patient-level representations, but rarely separate these two levels explicitly within a unified framework. \\

% What this paper adds &
% This paper introduces a framework that separates cohort-level typical pathway identification from patient-level inference. The approach provides interpretable typical pathways together with low-dimensional patient representations, allowing patients to be characterized as probabilistic mixtures of typical pathways rather than being assigned to a single pathway component. \\

% Who would benefit from the new knowledge in this paper &
% The proposed framework may support researchers and clinicians working with routine care data by providing an interpretable approach for summarizing heterogeneous healthcare trajectories, exploring patient subgroups, and studying variation in real-world clinical practice. \\

% \hline
% \end{tabularx}
% \end{table}

\section{Clinical data description}\label{data}

To illustrate the methodological challenges addressed in this work, we consider routinely collected inpatient coding data from prostate cancer patients undergoing radical prostatectomy at the Department of Urology, Medical Center-University of Freiburg. The data were processed within the framework of the German Medical Informatics Initiative \cite{lawal2016clinical,semler2018mii}. 
Hospital admission diagnoses were recorded using ICD-10 codes (International Statistical Classification of Diseases and Related Health Problems), treatment procedures were documented using OPS codes (Operation and Procedure Classification System, the German modification of the International Classification of Procedures in Medicine), and discharge was included as an additional event marking the end of the inpatient trajectory. Each event was linked to a patient identifier and timestamp, yielding a temporally ordered care trajectory for each patient. The left panel of Figure~\ref{fig:workflow} illustrates the structure of the routine care data, including temporally ordered clinical events, heterogeneous patient trajectories, and key characteristics of the cohort.

The dataset comprises 995 patients undergoing radical prostatectomy between 2015 and 2020, corresponding to 6,904 recorded inpatient clinical events. In total, the dataset contains 4 distinct ICD-10 admission diagnosis codes, 162 OPS procedure codes, and discharge information. The ICD-10 codes included C61 (malignant neoplasm of the prostate), I89.8 (other specified noninfective disorders of lymphatic vessels and lymph nodes), N13.3 (other and unspecified hydronephrosis), and C79.5 (secondary malignant neoplasm of bone and bone marrow). OPS procedures spanned multiple catalog chapters, primarily Chapter 5 (surgical procedures; 56 codes), followed by Chapters 3 (imaging diagnostics; 40 codes), 8 (non-operative therapeutic procedures; 37 codes), 1 (diagnostic measures; 22 codes), 9 (supplementary procedures; 6 codes), and 6 (medication-related procedures; 1 code).

Each clinical event was mapped to a numeric state representation to facilitate compact trajectory encoding and prefix-tree construction, resulting in 167 distinct state labels. State 0 corresponds to ICD-10 code C61, representing the primary admission diagnosis, while state 166 represents discharge.

Substantial structural heterogeneity is apparent at the cohort level. Patient trajectories vary considerably in length and composition: some patients undergo in-house fusion biopsy prior to surgery, whereas others are referred with externally completed diagnostics and therefore exhibit shorter trajectories. In addition, multiple events may occur at the same timestamp. For example, robotic-assisted radical prostatectomy is simultaneously coded with OPS-5.604 and OPS-5.98, producing partially overlapping event sets across patients. Although the underlying clinical processes are relatively standardized, variations in documentation timing, coding practices, and preparatory procedures generate a fragmented and highly branched cohort-level representation, resulting in numerous distinct pathway variants despite similar overall care structures.

These characteristics motivated the development of a structured framework that captures typical cohort-level care patterns while reducing patient-specific variability and preserving clinically meaningful transitions.

\section{Methodological framework towards subgroup identification} \label{Method}
\begin{figure}[ht]
    \centering
    %\hspace*{-0.5cm}
    \includegraphics[width=1\linewidth]{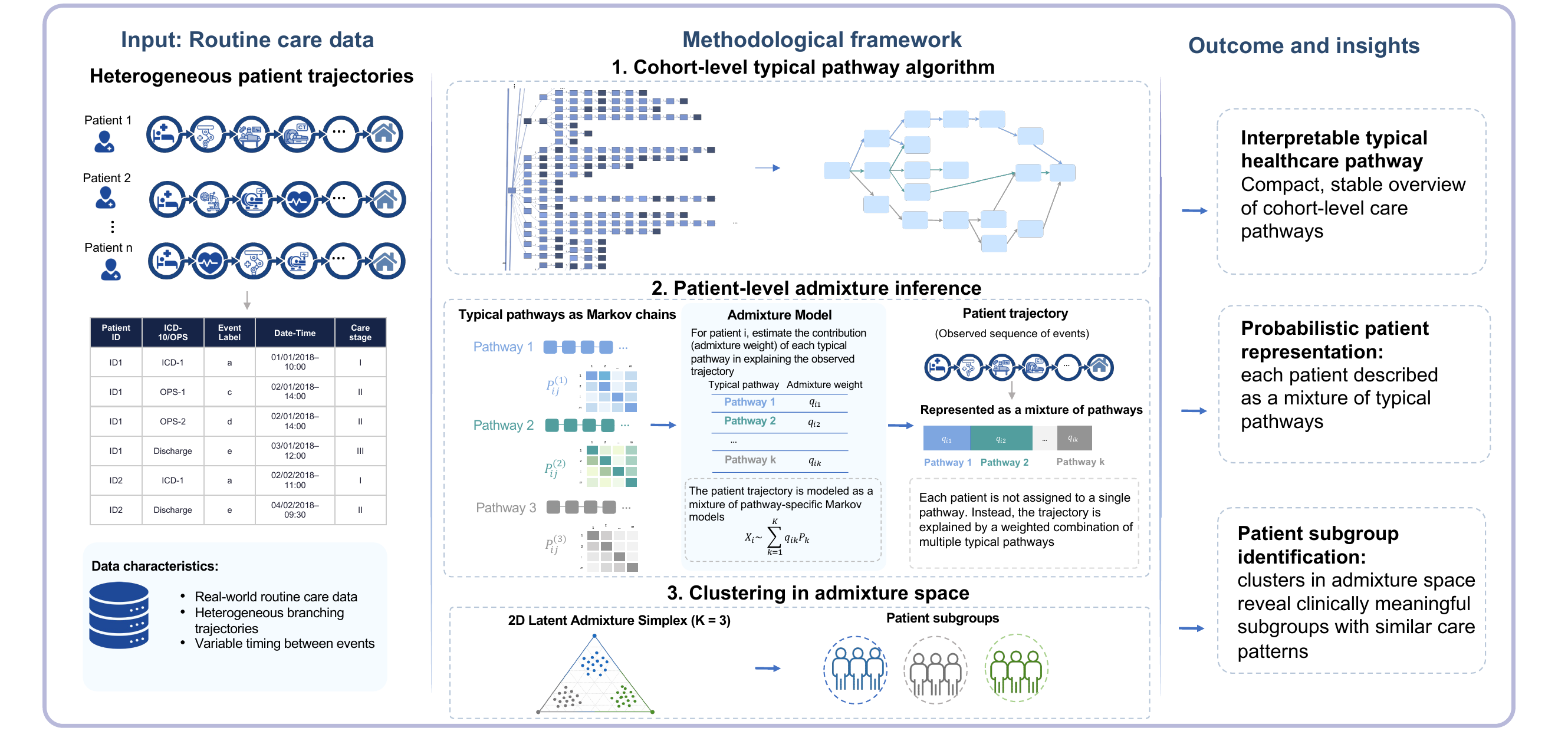}
    \caption{Overview of the proposed three-step framework for identifying typical healthcare pathways and modeling patient trajectories. Starting from heterogeneous routine care trajectories, the framework first derives cohort-level typical pathways through graph simplification strategies. These pathways are represented as pathway-specific Markov models and used in an admixture framework to infer patient-specific pathway contributions. The resulting admixture representations provide a low-dimensional summary of patient trajectories and support downstream subgroup identification in admixture space. The workflow also illustrates the event-level structure of the routine care data and key characteristics of the clinical trajectories.}
    \label{fig:workflow}
\end{figure}

Figure~\ref{fig:workflow} outlines our proposed methodological framework for analyzing heterogeneous routine-care trajectories by separating cohort-level pathway identification from patient-level inference. Using longitudinal clinical event data with branching care patterns, the approach first derives typical healthcare pathways that summarize common progression structures across the cohort (Step~1). It then applies an admixture-based modeling strategy to estimate the extent to which each individual trajectory is explained by multiple typical pathways, rather than a single fixed pattern (Step~2). These patient-level admixture representations support interpretable subgroup identification and subsequent downstream analyses (Step~3). A detailed description of each step in the methodological framework is provided in the following sections. To further evaluate the framework under controlled conditions, we additionally conduct a simulation study assessing its ability to recover latent pathway structures and patient-level admixture proportions under predefined structural mixing scenarios. Details of the simulation design and results are provided in Supplementary Section~S1. To \revtwo{4}{ensure}{assess} the stability of the identified structures and prevent information leakage, we adopt a strict training-test separation at the patient level prior to any modeling steps. In each repetition, patients are randomly partitioned without replacement into a training set comprising 80\% of the cohort and a test set comprising the remaining 20\%. The cohort-level pathway graph described in Step~1 is constructed exclusively using the training data. By deriving the pathways from this subset, we ensure that the "typical" care pathways serve as a fixed scaffold for the independent inference of patient-level admixture weights in Step~2. 
% \revtwo{4}{}{Repeated random patient-level train–test splits were used to assess sampling stability within the observed cohort. These splits evaluate sensitivity to the selection of training patients from the same center and observation period; they do not assess generalization across time periods, institutions, or clinical workflows.}

\subsection{Step 1: Cohort-level typical pathway algorithm} \label{Step1}

\begin{figure}[t]
        \centering
        \includegraphics[width=1\linewidth]{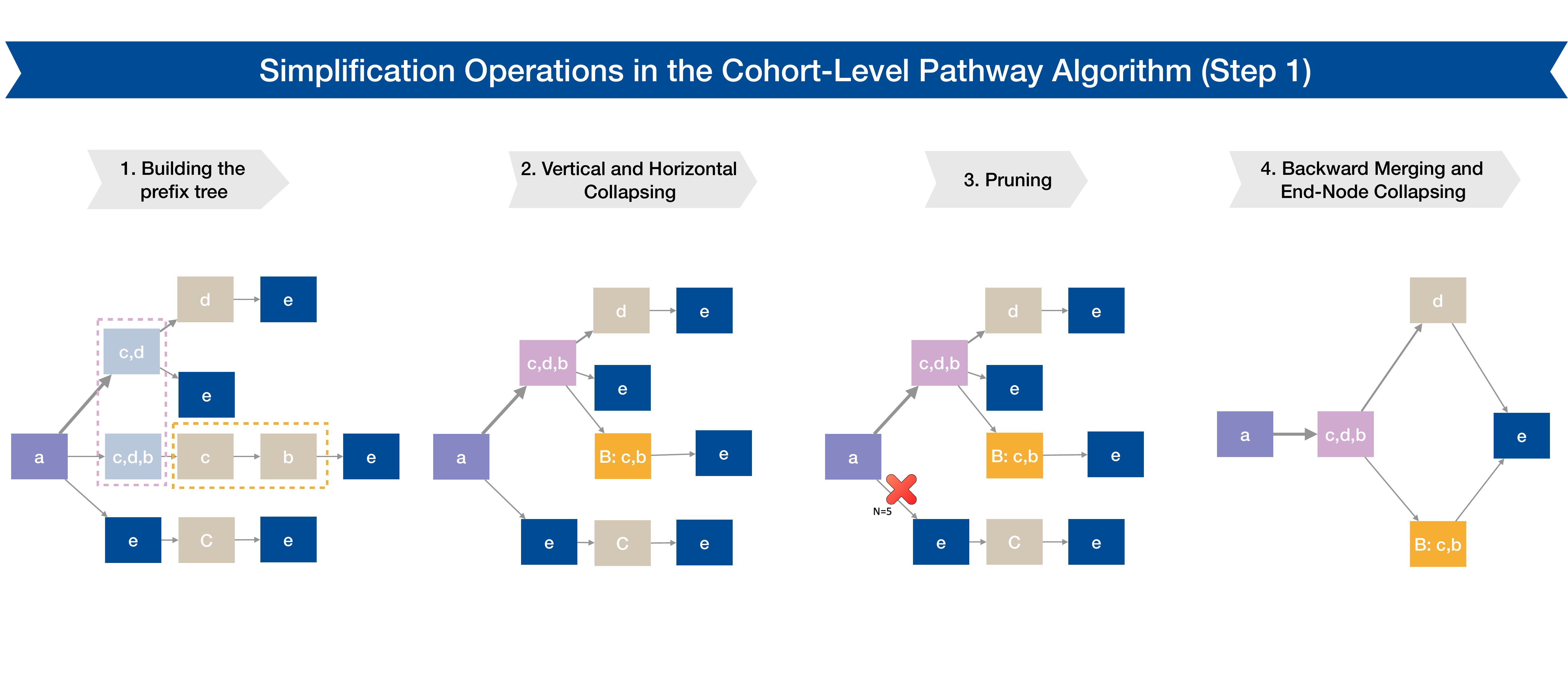}
        \caption{Illustration of the graph simplification operations used in the cohort-level pathway algorithm (Step 1). Vertical collapsing merges nodes with overlapping states within the same care stage, horizontal collapsing reduces sequences of less critical nodes, pruning removes infrequent states, and backward merging with end collapsing unifies redundant terminal nodes. Together, these operations transform raw prefix trees into compact and interpretable pathway graphs.
}
        \label{fig:simplification}
    \end{figure}

The cohort-level pathway algorithm constructs a simplified and interpretable pathway graph from temporally ordered patient trajectories through tree construction, collapsing operations, and pruning (Figure~\ref{fig:simplification}).

\subsubsection{Tree construction from raw data}
Each patient trajectory is defined as a temporally ordered sequence of event blocks. An event block denotes the set of clinical codes recorded for a patient at the same timestamp. We refer to these event blocks as \revtwo{5}{care stages, representing discrete steps in the patient trajectory corresponding to diagnostic, therapeutic, or administrative events, such as biopsy, surgical intervention, or discharge.}{observed coded care stages, representing discrete steps in the recorded patient trajectory corresponding to diagnostic, therapeutic, or administrative events. Grouping codes recorded at the same timestamp avoids imposing an ordering between events for which no finer temporal ordering is available in the data. However, simultaneous recording does not necessarily imply that the corresponding events constitute a distinct underlying clinical state, as code bundles may also reflect documentation and coding conventions.}

\revtwo{5}{}{Discharge is treated as an observed administrative terminal event
marking the end of the inpatient trajectory. It is part of the
coded trajectory representation and is not used as an external
outcome label. More generally, no external clinical outcome or
predefined patient-group label is used to construct the pathway
graph, estimate patient-specific admixture weights, or perform
clustering in admixture space.}
The prefix tree is constructed iteratively by inserting patient trajectories one at a time. For the first patient, the ordered sequence of event blocks is represented as a path of nodes connected by directed edges, starting from a common root node and extending stepwise until discharge. For each subsequent patient, the trajectory is inserted sequentially from the root. At each step, the current event block is compared with the existing child nodes of the current node. If a child node with the same set of clinical events already exists, the patient is assigned to that node and its count is incremented. If no matching child node exists, a new node is created and attached as a child of the current node. This procedure is repeated until the full trajectory has been processed. In this way, patients sharing identical initial segments follow the same path in the tree, while differences at later stages give rise to new branches. If a patient’s first event block differs from existing ones, a new branch is created directly from the root.
Formally, each node \(N_i\) is characterized by:
\[
N_i = (s_i, c_i, \mathcal{I}_i),
\]
where \(s_i\) denotes the set of clinical events at the node, \(c_i\) the number of patients passing through it, and \(\mathcal{I}_i\) the corresponding patient identifiers. Directed edges connect consecutive nodes in temporal order, and edge weights correspond to transition frequencies. 

\subsubsection{Node and procedure importance} \label{Node importance}

To identify structurally informative nodes of the prefix tree, we assign an importance 
score based on their relative frequency within each care stage. The importance score is designed to highlight nodes that distinguish between alternative patient trajectories. Nodes that occur in an intermediate proportion of patients within a care stage are considered most informative, because they split the cohort into comparably sized subgroups. In contrast, nodes that occur only rarely represent isolated deviations, whereas nodes that occur in nearly all patients correspond to routine events that provide little structural differentiation. We say $N_i \sim N_j$ 
if nodes $N_i$ and $N_j$ belong to the same care stage, and define
\begin{equation}
    \mathcal{C}_i := \{j : N_j \sim N_i\}
\end{equation}
as the index set of all nodes in the same care stage as $N_i$. Care stages form disjoint partitions of the nodes, such that each node belongs to exactly one care stage. Consequently, if \(j \in \mathcal{C}_i\), then \(\mathcal{C}_j = \mathcal{C}_i\). Furthermore, \(\sum_{j \in \mathcal{C}_i} c_j\) represents the total number of patients assigned to nodes within the corresponding care stage. Within a given care stage, each patient contributes to at most one node, although patients may contribute to nodes across multiple care stages throughout their trajectory.

The frequency of node $N_i$ relative to the other nodes in the same care stage is:
\begin{equation}
    p_i = \frac{c_i}{\sum_{j \in \mathcal{C}_i} c_j}.
\end{equation}
The importance of node $N_i$ is then defined as:
\begin{equation}
    \text{Imp}(N_i) = \min(p_i,\, 1-p_i) \cdot \sum_{j \in \mathcal{C}_i} c_j.
\end{equation}

This formulation assigns maximal importance when the relative frequency $p_i$ is close to 0.5, corresponding to nodes that divide the cohort into two comparably sized subgroups at that stage. As $p_i$ approaches $0$ or $1$, the term $\min(p_i, 1-p_i)$ decreases symmetrically, reflecting that such nodes correspond either to rare deviations or to routine events shared by nearly all patients. Multiplying by the total stage size 
$\sum_{j \in \mathcal{C}_i} c_j$ ensures that nodes occurring in larger stages receive proportionally greater weight, thereby favoring structurally informative nodes supported by a substantial number of patients.

The importance of a procedure $\mathcal{P}_k$, which may occur in multiple nodes, is defined as the sum of importance scores over all nodes in which it appears:
\begin{equation}
    \text{Imp}\mathcal{P}_k = \sum_{i: \pi_k \in s_i} \text{Imp}(N_i),
\end{equation}
where \(s_i\) denotes the set of clinical states at node \(N_i\). This aggregation reflects the idea that a procedure is structurally important if it repeatedly appears in nodes that contribute to meaningful branching of patient trajectories.

Importance scores are computed for all nodes and procedures in the tree. Those exceeding a predefined threshold are designated as important and used to guide subsequent collapsing and pruning steps. The importance threshold is specified by the user and controls the structural resolution of the resulting pathway graph. Lower threshold values retain a larger number of nodes and preserve finer-grained variations, whereas higher values emphasize the most recurrent and structurally central transitions. In practice, the threshold therefore regulates the trade-off between structural detail and interpretability of the resulting pathway graph. The resulting importance scores are used to guide the simplification of the prefix tree. Such simplification is necessary because a raw prefix tree may retain nearly all patient-level variation, including rare branches, minor coding differences, concurrent event combinations, and documentation-specific deviations. While this detailed representation remains faithful to the original event data, it can obscure dominant cohort-level care structures and reduce the interpretability of the resulting graph. In the following, we outline collapsing and pruning operations that reduce structural complexity while preserving clinically meaningful progression patterns.

\subsubsection{Vertical collapsing}

Vertical collapsing simplifies the prefix tree by merging nodes within the same care stage that share an important procedural subset. If two nodes both contain an identified important procedure (or procedure combination), they are interpreted as representing the same core structural event, even if they differ in additional codes.

Formally, consider two nodes
\[
N_i = (s_i, c_i, \mathcal{I}_i)
\quad \text{and} \quad
N_j = (s_j, c_j, \mathcal{I}_j),
\]
with \(i < j\),
belonging to the same care stage. If there exists a non-empty subset of clinical codes \(X\) such that \(X \subseteq s_i\), \(X \subseteq s_j\), and every procedure \(P_k \in X\) satisfies \(\mathrm{Imp}(P_k) \geq \tau\) as defined in Section~\ref{Node importance}, then the shared subset is interpreted as representing a structurally important event pattern. Because procedure importance aggregates relative node frequencies \(p_i\) across care stages, this criterion favors merging nodes associated with informative branching structures rather than rare or near-universal events, whereby node \(N_j\) is merged into \(N_i\) and the attributes of \(N_i\) are updated as
\[
s_i \leftarrow s_i \cup s_j, 
\qquad
c_i \leftarrow c_i + c_j,
\qquad
\mathcal{I}_i \leftarrow \mathcal{I}_i \cup \mathcal{I}_j.
\]
Node $N_j$ is then removed from the tree, and all incoming and outgoing edges previously connected to $N_j$ are redirected to $N_i$.

After merging $N_j$ into $N_i$, each child node of $N_j$ is reattached as a child of the updated node $N_i$. The existing children of $N_i$ are preserved. The vertical collapsing condition is then re-evaluated among all children of $N_i$. If any child nodes share a non-empty subset of clinical codes whose procedures satisfy the predefined importance threshold, they are merged accordingly. This procedure is applied recursively along descendant branches until no further vertical merging is possible.

For example, as illustrated schematically in Panel~2 of Figure~\ref{fig:simplification}, if the procedure combination $[c,d]$ is identified as important within a care stage, nodes representing $[c,d]$ and $[c,d,b]$ are merged by updating one node and removing the other. The children of the removed node are reattached to the merged node, and any children satisfying the vertical collapsing condition are subsequently merged. This recursive vertical collapsing consolidates parallel branches within a stage that share important clinical content, thereby reducing structural redundancy while preserving differentiation across healthcare pathways.

\subsubsection{Horizontal collapsing}
Horizontal collapsing, also referred to as longitudinal path collapsing, reduces the number of intermediate nodes along patient trajectories by merging consecutive nodes that are not designated as important, i.e., whose importance scores fall below the predefined threshold. Sequences of adjacent non-important nodes along a single path, located between two important nodes or between an important node and a branching point, are collapsed into a single composite node.

Formally, consider a consecutive chain of nodes %$N_{i_1}, N_{i_2}, \ldots, N_{i_k}$ along a single path in the tree, 
\(N_{i_1}, N_{i_2}, \ldots, N_{i_k}\) such that \(N_{i_{j+1}}\) is a child of \(N_{i_j}\) for \(j = 1, \ldots, k-1\), where none of the nodes is designated as important. Since these nodes lie along a single trajectory segment, they represent the same set of patients, i.e.,
\[
\mathcal{I}_{i_1} = \mathcal{I}_{i_2} = \cdots = \mathcal{I}_{i_k},
\]
and therefore share the same patient count.

This chain is merged by updating the first node in the sequence:
\[
N_{i_1} \leftarrow 
\left(
\bigcup_{j=1}^{k} s_{i_j},
\; c_{i_1},
\; \mathcal{I}_{i_1}
\right),
\]
and removing the nodes $N_{i_2}, \ldots, N_{i_k}$ from the tree. The updated node is connected directly to the nearest preceding and succeeding retained nodes.

For example, as shown in Panel~2 of Figure~\ref{fig:simplification}, if procedures $c$ and $b$ are not designated as important and occur consecutively along the same branch, they are merged into a single composite node (e.g., $B:\{c,d\}$), representing the combined set of states until the next important procedure is reached (here, $e$). This operation reduces structural complexity by compressing consecutive non-important nodes while preserving transitions between important clinical states.

\subsubsection{Pruning}
After the collapsing operations, pruning is applied to further refine the pathway graph by removing nodes that correspond to infrequent or weakly supported variations. This step reduces minor branches while retaining transitions that are sufficiently supported in the cohort. A node $N_i$ is retained only if it satisfies both an absolute and a relative support criterion. Formally, let $N_p$ denote the immediate parent of node $N_i$. A node is preserved if
\[
c_i \geq \alpha 
\quad \text{and} \quad 
\frac{c_i}{c_p} \geq \beta,
\]
where $c_i$ denotes the number of patients passing through node $N_i$, and $c_p$ denotes the patient count of its parent node $N_p$. The absolute threshold $\alpha$ removes nodes observed in few patients, whereas the relative threshold $\beta$ eliminates nodes that retain only a small proportion of patients relative to their parent node; nodes representing only a small continuation of patients from the preceding step are removed.

%The specific parameter values used for pruning are reported in the Results section.

\subsubsection{Backward merging and end-node collapsing}

Vertical collapsing is performed in a forward (top-down) manner from the initial node towards terminal nodes. At each care stage, nodes are merged if they satisfy the importance-based merging criterion. However, merging may fail at an intermediate stage even though later stages again satisfy the criterion. Because forward collapsing propagates only downward, such merge opportunities may remain unresolved. To address this, a backward merging step is applied. Starting from terminal nodes (nodes without outgoing edges, often corresponding to discharge or other final outcome states), the algorithm recursively traverses upward and re-evaluates the same importance-based merging criterion used in vertical collapsing. If nodes satisfy the criterion, they are merged using the same in-place update logic described earlier; otherwise, existing parent--child relationships are preserved. In contrast to forward collapsing, where child nodes are consolidated under a merged parent, backward merging consolidates parent nodes. After merging, parent nodes are reattached to the updated node and recursively re-evaluated until no further merging is possible. This bottom-up pass consolidates structurally equivalent nodes that were not merged during the initial forward traversal.

%%%%%%%%%%%%%%%%

\subsection{Step 2: Patient-level admixture inference} \label{Step2}
To capture patient-level heterogeneity, we employ a mixed-membership (admixture) model in which each patient's observed trajectory is represented as a probabilistic combination of the pathway-specific Markov chains derived from the Step 1 typical pathways. Thus, the model does not assign each patient to exactly one pathway. Instead, it estimates admixture weights that quantify the extent to which each trajectory is explained by each cohort-level pathway component.

\subsubsection{Pathway-specific Markov models and admixture formulation}

\revone{1}{Using the cohort-level typical pathway graph derived from the training set, we identify $K$ typical pathways that represent the main care progression patterns observed in the cohort. Each pathway represents an ordered sequence of nodes in the simplified graph, starting from the root node (admission) and ending at a terminal node (typically discharge).
For each pathway \(k\), we construct a discrete-time, first-order Markov chain, yielding a collection of \(K\) pathway-specific Markov models defined over a common state space. The transition probability matrix $P_k$ is estimated from empirical transition frequencies along the subgraph associated with pathway $k$. Each entry
$
P_k(i,j)
$
denotes the probability of transitioning from state $i$ to state $j$ within pathway $k$. Formally, transition probabilities are estimated as
$$
P_k(i,j)
=
\frac{n_k(i,j)}
{\sum_{j'} n_k(i,j')}
$$
where \(n_k(i,j)\) denotes the empirical number of observed transitions from state \(i\) to state \(j\) within pathway \(k\). Thus, probabilities sum to one for each originating state, and transitions not observed within pathway \(k\) are assigned probability zero.}{ In the simplified pathway graph derived from the training set, the root node represents admission. The nodes directly connected to the root represent the different first care steps after admission; in our application these were $(106,119)$, $(106)$, and $(10,119)$ (Figure~3). We call each of these nodes, together with all nodes that follow it in the graph, a \emph{pathway branch}, and we define one pathway component per branch. The number of components $K$ is therefore the number of nodes directly connected to the root after simplification. It is not chosen manually but follows from the importance threshold and the collapsing and pruning rules of Section~3.1. A branch may contain several root-to-terminal paths (for example, after a readmission); all of them belong to the same component.
For each component $k$, we construct a discrete-time, first-order Markov chain over a common state space. The transition probability matrix $P_k$ is estimated from the training trajectories of the patients in branch $k$, i.e., the patients whose first care step after admission is the first node of that branch, including patients whose trajectory segments were merged into it during simplification. \\
Formally, 
\begin{equation}
P_k(i,j)
=
\frac{n_k(i,j)}
{\sum_{j'} n_k(i,j')}
\end{equation} 
where $n_k(i,j)$ is the number of transitions from state $i$ to state $j$ in these trajectories. Thus, probabilities sum to one for each originating state, and transitions not observed for component $k$ are assigned probability zero. Patients whose first care step was removed by pruning do not contribute to any $P_k$, but admixture weights are estimated for all patients.}
Because all pathway models are defined over the same state space, the resulting collection \(\{P_1,\dots,P_K\}\) yields directly comparable pathway-specific transition structures.

Care progression within each pathway is modeled under a first-order Markov assumption, whereby the probability of transitioning to the next state depends only on the current state. Although real-world healthcare trajectories may exhibit longer-range dependencies, the simplified pathway representation is intended to capture dominant structural transitions, for which a first-order model provides a parsimonious and stable approximation.
For a patient $i$ in the test set (used to evaluate pathway stability and avoid information leakage), let
$
q_i = (q_{i1}, \dots, q_{iK})
$
denote the patient-specific admixture weights over the \(K\) pathway components, with $q_{ik} \ge 0$ for all $k$ and $\sum_{k=1}^{K} q_{ik} = 1$. Let $S_i = (s_{i1}, \dots, s_{iT_i})$ denote the observed trajectory of patient $i$. 

Under the admixture model, the probability of each observed transition is represented as a weighted combination of the corresponding pathway-specific transition probabilities. The log-likelihood is therefore given by
\[
\log p(S_i \mid q_i)
=
\sum_{t=1}^{T_i-1}
\log \left(
\sum_{k=1}^{K}
q_{ik}\, P_k(s_{it}, s_{i,t+1})
\right),
\]
where $(s_{it}, s_{i,t+1})$ denotes consecutive states.

The admixture weights $q_i$ are inferred by maximizing this log-likelihood subject to the simplex constraint
\[
q_i \in \Delta^{K-1} = \{ q_i \in \mathbb{R}^K : q_{ik} \ge 0,\ \sum_{k=1}^{K} q_{ik} = 1 \},
\]
while the pathway-specific transition matrices $P_k$ remain fixed as estimated from the training data.
\revone{2}{}{For numerical evaluation of the log-likelihood, each mixture
transition probability was bounded below by $\varepsilon=10^{-8}$
before taking its logarithm. A transition absent from one pathway
component could still receive positive probability from another
component with positive admixture weight. Transitions unsupported
by all components contributed the constant $\log(\varepsilon)$,
independently of the patient's admixture weights. This numerical
floor was applied during likelihood evaluation without modifying
or renormalizing the pathway-specific transition matrices.
The reported values therefore represent numerically floored
log-likelihood scores.}
Two optimization strategies were used to estimate the admixture weights: an expectation–maximization (EM) procedure and a constrained numerical optimization using sequential least squares programming (SLSQP). Both approaches estimate the admixture weights \(q_i\) while keeping the pathway-specific transition matrices \(P_k\) fixed. SLSQP performs a direct constrained optimization of the observed-data log-likelihood under the simplex constraint on \(q_i\). In contrast, the EM procedure introduces latent pathway assignments for individual transitions and alternates between estimating these latent assignments and updating the admixture weights, thereby iteratively increasing the observed-data log-likelihood until convergence. The optimization was initialized with uniform admixture weights and iterated until convergence of the log-likelihood.

The inferred admixture vectors provide a low-dimensional representation of patient trajectories in terms of contributions from the $K$ pathway-specific models. This representation maps heterogeneous trajectories of varying length and composition to a common latent space and forms the basis for subsequent subgroup analysis.

\subsection{Step 3: Clustering in admixture space} \label{step3}
The inferred admixture vectors \(\{q_i\}\) define a shared low-dimensional representation of patient trajectories. Each patient is represented as a point in the same \(K\)-dimensional simplex, enabling direct comparison across trajectories despite differences in trajectory length and heterogeneous event composition. This representation removes the need to align individual event sequences before comparison. In the proposed framework, patient trajectories are clustered based on their inferred admixture vectors using \(k\)-means clustering with Euclidean distance in the admixture space. The use of \(k\)-means provides a simple and interpretable partitioning of patients according to similarities in their estimated pathway composition. This step is intended to identify patient groups with similar mixtures of pathway-specific patterns, rather than to recover complex or non-convex cluster structures. \revtwo{3}{The number of clusters is set to \(K\), corresponding to the number of pathway-specific models. This choice facilitates direct comparison between the inferred pathway structures and the patient groups obtained from clustering in admixture space.}{ We set the number of clusters equal to the number of pathway components, $K_c = K$, so that each cluster can be related to a dominant pathway composition. This choice was validated by comparing $K_c = 2,\ldots,6$ (Supplementary Section~S3.2).}

\revone{4}{}{\subsection{Comparison with alternative approaches}
We compared our framework with the approaches discussed in the Introduction. As our framework both identifies typical pathways at the cohort level and represents patients and subgroups at the patient level, while each of the other approaches addresses only one of the two tasks, we compared them separately for each level. All methods used the training patients of the same 20 train--test splits as the stability analysis and identical input data. \textit{Cohort level.} The typical-pathway graph was compared with the directly-follows graph (DFG) and the Inductive Miner (PM4Py 2.7.23.8), discovered from the raw training event log. As a DFG is always frequency-filtered in practice, it was filtered to the same number of transition types as the typical-pathway graph. We assessed the share of test transitions covered by the model, cross-split stability (edge Jaccard similarity), cycles, and nodes not reachable from the start. \\
\textit{Patient level.} The admixture representation was compared with latent Dirichlet allocation (LDA), mixtures of first-order Markov chains, mixtures of hidden Markov models, edit-distance clustering with $k$-medoids, and $k$-means on transition frequencies, all with $K = 3$. As an ablation, we used the same admixture model with components learned freely by EM instead of being taken from the Step~1 pathways. We assessed the stability of the patient clusters across training sets on complete trajectories (adjusted Rand index, ARI, and normalized mutual information, NMI).} \revone{3}{}{To relate the clusters to an external variable, we used a subsequent recorded admission. Because later admissions are part of the patient trajectories, they are already used to estimate the pathways and clusters, so the clusters cannot be compared with this outcome directly. We therefore truncated each trajectory at the first discharge, keeping only the events of the first hospital stay, refitted all methods on these truncated trajectories using the training patients only, and determined only afterwards whether a subsequent admission was recorded for the test patients. For each method, we assessed the association (Cram\'er's~$V$) between its test-patient clusters and a subsequent admission. For our framework, we additionally report the proportion of patients with a subsequent admission in each cluster with 95\% Wilson confidence intervals, an overall Fisher exact test, and pairwise Fisher exact tests with Holm adjustment for multiple testing (results in Section~\ref{4.4}).}

%%%%%%%%%%%%%%%%%%%%%%%%%%%%%%%%%%%%%%%%%%%%%%%%%

\section{Application of methodological framework to clinical data} \label{Result}
In the following, we apply the framework to the prostate cancer cohort (Section~\ref{data}) and summarize the results.

\subsection{Typical healthcare pathways in the prostate cancer cohort}
\begin{figure}[ht]
    \centering
    \includegraphics[width=1\linewidth]{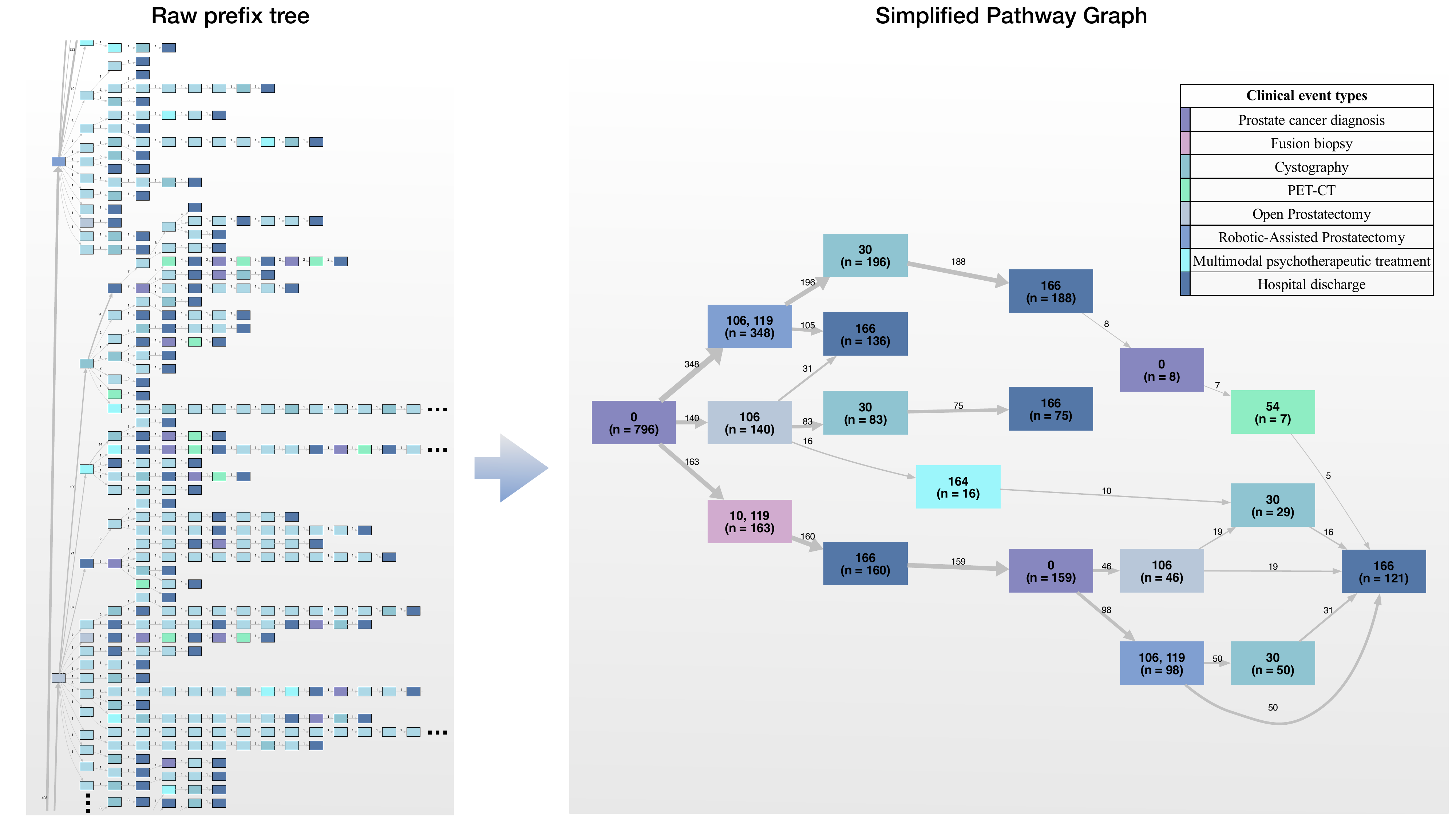}
    \caption{Left: prefix tree representation of raw patient trajectories, illustrating pronounced heterogeneity and extensive branching. Right: simplified pathway graph highlighting typical healthcare pathways in the prostate cancer cohort after collapsing and pruning. The pathway graph is constructed exclusively from the training data of one representative train–test split, while test data are not used in this step. Edge thickness reflects transition frequency. %Pathway graphs and pathway-specific Markov chain models used for patient-level inference are derived from training data only
    }
    \label{fig:real_data}
\end{figure}

The cohort-level typical pathway identification step first yielded a prefix tree representation capturing shared and diverging trajectory segments across patients (Figure~\ref{fig:real_data}, left). While common progression segments were visible in early stages, trajectories increasingly diverged in later stages, resulting in extensive branching and structural fragmentation.

After applying our typical pathway algorithm, the simplified pathway graph (Figure~\ref{fig:real_data}, right) revealed consolidated cohort-level structures that were not clearly discernible in the raw tree representation. Nodes retained after importance-based filtering primarily corresponded to structurally central diagnostic and procedural steps, whereas removed nodes mainly reflected infrequent variations. The resulting graph provides a compact and interpretable representation of typical care structures while preserving the overall progression from admission to discharge. A total of 29.8\% of the original patient trajectories were fully retained in the final simplified pathway graph, meaning that their complete trajectories remained represented despite the collapsing and pruning procedures.  

Clinically, the graph reflects typical pathways observed in prostate cancer patients within the cohort. A clear backbone progression is visible from diagnostic evaluation to biopsy confirmation, followed by surgical treatment and discharge, corresponding to key procedural stages consistently observed across most patient trajectories. At the same time, the graph captures smaller side branches representing variation in diagnostic workup. Some patients underwent additional diagnostic coding events prior to surgery, whereas others transitioned more directly from diagnosis to operative treatment. These differences likely reflect variation in referral patterns, such as externally completed diagnostics versus more extensive in-hospital diagnostic workup.
Node and procedure importance scores were strongly right-skewed, with few high-scoring elements and many low-scoring ones. To assess the sensitivity of the framework to parameter selection, we varied the importance and pruning thresholds and observed that these changes primarily affected peripheral branches and graph resolution, while the typical pathways remained stable across parameter settings (Supplementary Figure~S3). For the main analysis, we used \(\tau = 15\), \(\alpha = 5\), and \(\beta = 0.1\). \revone{5}{Across multiple random train--test splits, the same three pathway structures were consistently recovered, with only minor differences in low-support branches (Supplementary Figure~S2).}{In an additional analysis of 20 patient-level train--test splits, all three main initial pathway branches were recovered in every split. At $\tau = 15$, pairwise directed-edge Jaccard similarity between the displayed simplified pathways had a median of 0.882 (IQR: 0.833--0.944; minimum: 0.684). These results support consistent recovery of the main initial branches and substantial overlap of the displayed transitions, with some variation in downstream structure (Supplementary Section~S2.1).}

\begin{figure}[ht]
    \centering
    \includegraphics[width=1\linewidth]{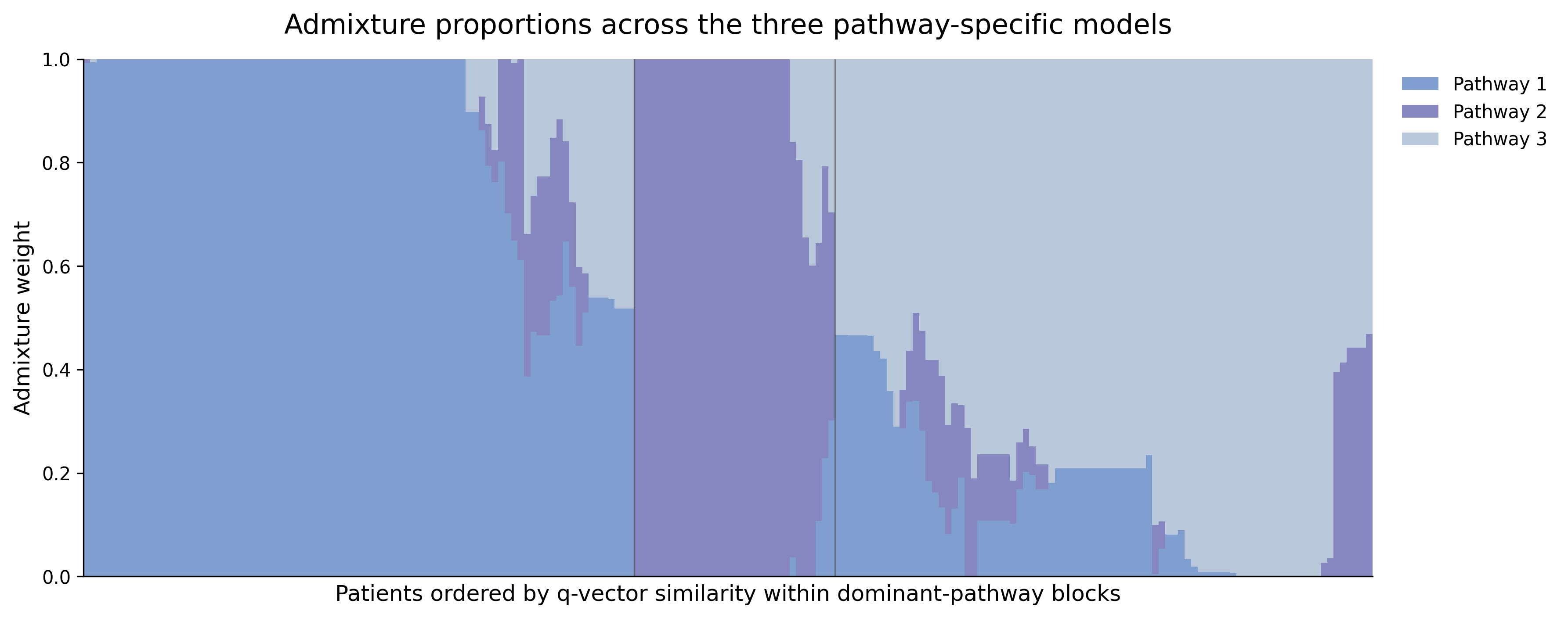}
    \caption{Stacked bar plots showing admixture proportions across the $K = 3$ inferred pathway models, with each bar corresponding to a patient in one representative test set (approximately 20\%  of the cohort, $n = 199$) from a single train–test split. Admixture weights are inferred only for test patients, while pathway models are estimated on the corresponding training set.}
    \label{fig:admixture_clustering}
\end{figure}

% Step 2: Admixture
\subsection{Admixture representations of patient trajectories}
\revone{1}{Three pathway-specific Markov models were derived from the typical pathways identified in the simplified graph.}{ The simplified graph had three branches starting at the root, (106, 119), (106) and (10, 119), from which three pathway-specific Markov models were derived. The same three branches were obtained in all 20 train–test splits and for all importance thresholds between 5 and 30 (Supplementary Section S3.1).} The upper pathway is characterized mainly by progression from the combined diagnostic and procedural state \((106,119)\) toward intervention \((30)\) and discharge \((166)\), with additional readmission-related transitions. The middle pathway follows a similar overall progression but includes further intermediate stages before discharge. The lower pathway reflects trajectories in which patients first undergo diagnosis and biopsy-related procedures \((10,119)\), are discharged \((166)\), and subsequently return through readmission \((0)\) for additional diagnostic or treatment-related procedures, including \((106,119)\) and intervention \((30)\), before a later discharge.

Based on these pathway-specific models, admixture representations were inferred for patients in the test sets. Each patient trajectory was summarized by a three-dimensional admixture vector, whose components quantify the relative contribution of the three typical pathway models. In this way, heterogeneous event sequences are represented in a common low-dimensional space while preserving information on pathway composition.

The admixture profiles varied across patients. Many trajectories were dominated by a single pathway model, whereas others showed mixed contributions, suggesting blended or intermediate care trajectories (Figure~\ref{fig:admixture_clustering}). This pattern is also reflected in two-dimensional projections of the admixture space (Figure~\ref{fig:admixture}A), where patients form three groups corresponding to dominant pathway contributions. Points near the simplex corners represent trajectories mainly explained by one pathway model, while points in intermediate regions indicate mixed trajectories.

% Step 3: Clustering
\subsection{Patient cluster structure in admixture space}\label{4.3}
\begin{figure}[H]
    \centering
    \begin{subfigure}[t]{0.48\textwidth}
        \centering
        \includegraphics[width=\textwidth]{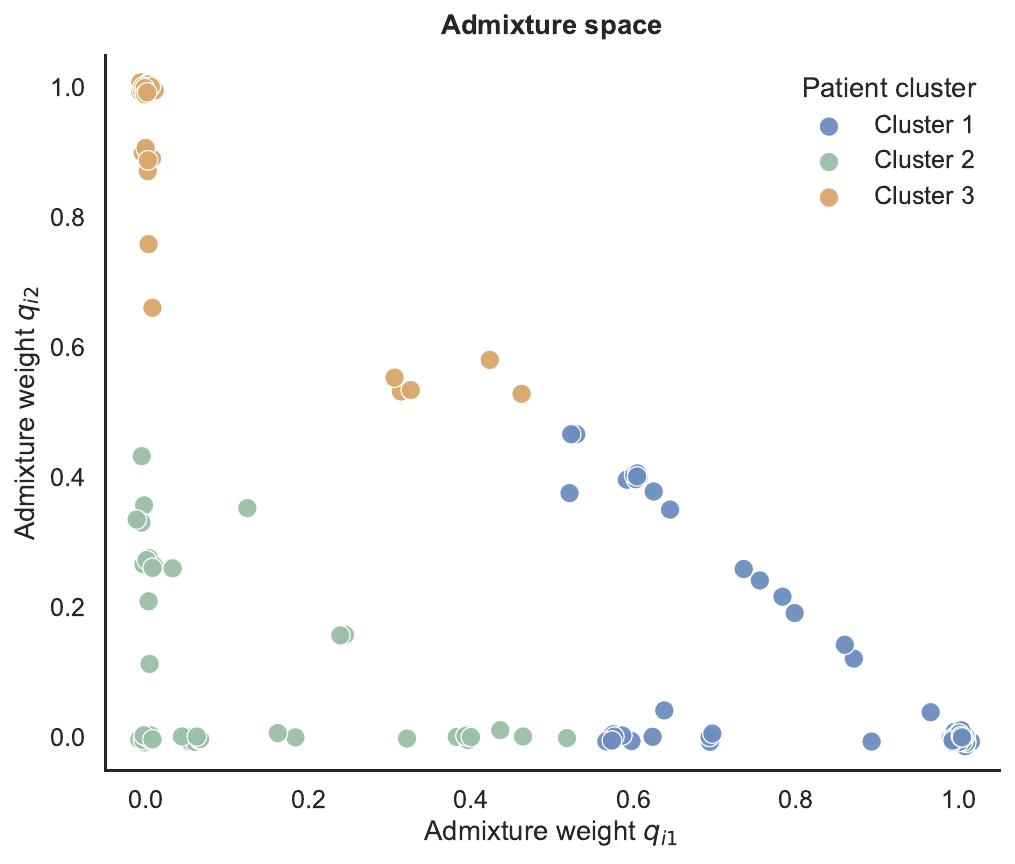}
        \caption{}
        \label{fig:q-scatter}
    \end{subfigure}
    \hfill
    \begin{subfigure}[t]{0.48\textwidth}
        \centering
        \includegraphics[width=\textwidth]{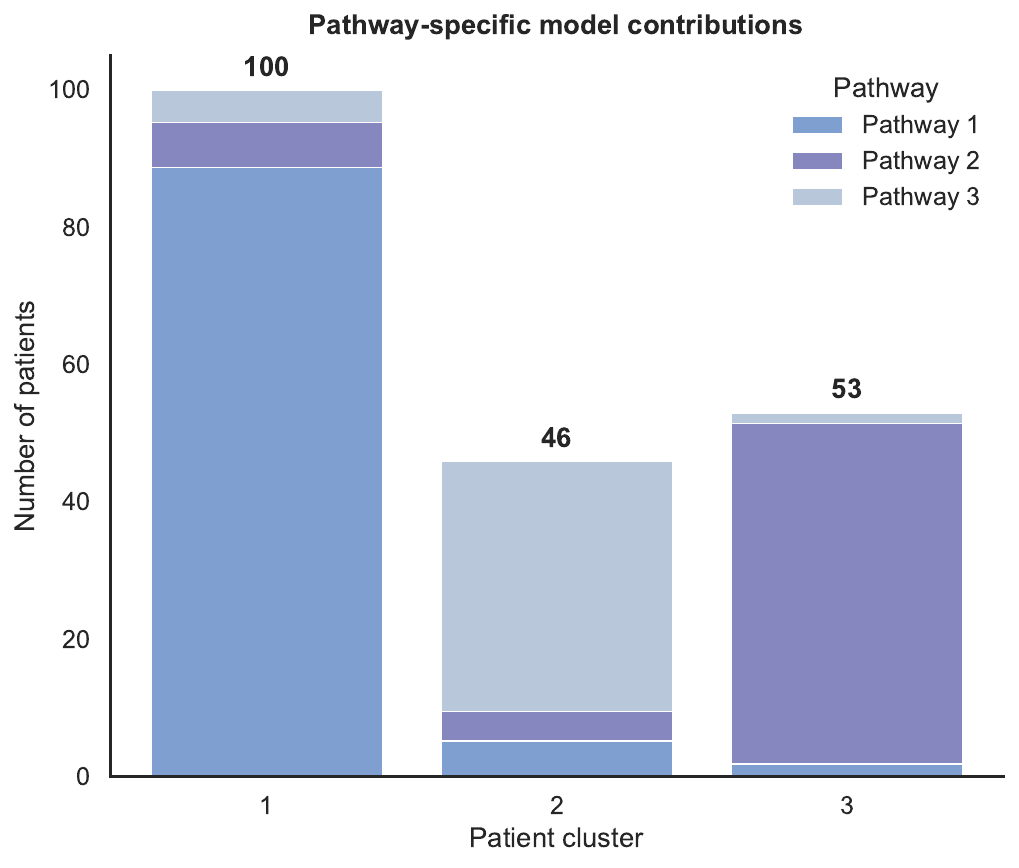}
        \caption{}
        \label{fig:cluster-bar}
    \end{subfigure}
    \caption{Admixture space and subgroup structures. (A) Two-dimensional projection of admixture vectors. (B) Cluster sizes with average admixture composition. Comparable results across additional splits are shown in the Supplementary Material.}
    \label{fig:admixture}
\end{figure}

Clustering performed on the admixture representations yielded well-separated patient groups characterized by distinct pathway compositions. Cluster sizes and average admixture compositions are shown in Figure~\ref{fig:admixture}.

\revone{5}{Across multiple random train--test splits, the same three typical pathway structures were consistently identified, with comparable admixture distributions and stable clustering results.}{Across 20 train--test splits, the patient clusters were reproducible (median adjusted Rand index between the clusterings of different splits 0.81 at $\tau = 15$; Supplementary Section~S2.1).}

Internal clustering validation measures computed on the admixture representations (Table~\ref{tab:cluster_validation}) further supported subgroup separation. For EM optimization, the Silhouette score averaged $0.70 \pm 0.03$, the Calinski--Harabasz index $562 \pm 64$, and the Davies--Bouldin index $0.44 \pm 0.07$. SLSQP optimization yielded consistently stronger separation, with a mean Silhouette score of $0.75 \pm 0.02$, a Calinski--Harabasz index of $727 \pm 50$, and a Davies--Bouldin index of $0.37 \pm 0.04$. Across the three train--test splits, the log-likelihood of the observed trajectories under the estimated admixture parameters was consistently higher for SLSQP than for EM ($-1808.51$ vs.\ $-1818.68$; $-1930.26$ vs.\ $-1940.92$; $-1965.37$ vs.\ $-1975.69$), indicating a better fit to the admixture objective and consistent improvement in cluster separation.
\revtwo{3}{}{When the number of clusters was varied from 2 to 6, $K_c = 3$ gave the lowest Davies--Bouldin index,
and larger $K_c$ mainly separated small groups of patients without improving the association with
subsequent recorded admission (assessed on trajectories truncated at the first discharge; Supplementary Section~S3.2).}
\begin{table}[ht]
\centering
\caption{Internal clustering validation indices computed on admixture representations across \revone{5}{three independent}{the three original} train--test splits. Results are reported as mean $\pm$ standard deviation.\revone{5}{}{ The 20-split stability analysis is reported in Supplementary Section~S2.1.}}
\label{tab:cluster_validation}
\begin{tabular}{l
                S[table-format=1.3(2)]
                S[table-format=3.0(2)]
                S[table-format=1.3(2)]}
\toprule
\textbf{Optimizer} &
\textbf{Silhouette} &
\textbf{Calinski--Harabasz} &
\textbf{Davies--Bouldin} \\
\midrule
EM    & 0.698 \pm 0.030 & 562 \pm 64 & 0.436 \pm 0.070 \\
SLSQP & 0.751 \pm 0.016 & 727 \pm 50 & 0.366 \pm 0.044 \\
\bottomrule
\end{tabular}
\end{table}

\revone{4}{}{\subsection{Comparison with alternative approaches} \label{4.4} \textit{Cohort level.} At the same size, the DFG retained largely the same clinical states as the typical-pathway graph, with comparable held-out coverage and stability (Table~2). However, it contained cycles in all splits, because every occurrence of an event is merged into one node. Index stays and readmissions are therefore mixed: in the typical-pathway graph, 183 of 184 patients discharged after an initial $(10,\,119)$ stay were readmitted, compared with 9 of 182 after $(106,\,119)$ and imaging, whereas in the DFG all readmissions pass through one edge (Supplementary Figure~S5).\\
\textit{Patient level.} The admixture clusters were more stable than those of all other methods (ARI 0.81 vs.\ 0.31--0.70; Table~3), including k-means on transition frequencies (0.62) and edit-distance clustering (0.70). Learning the components freely instead of using the Step~1 pathways reduced stability from 0.81 to 0.46.} \\

\revone{3}{}{\textit{Clinical association.} On trajectories truncated at the first discharge, each cluster of our framework was dominated by one of the three pathways, and the proportion of patients with a subsequent admission differed clearly between them: 94.2\% (65/69; 95\% CI 86.0--97.7\%) in the cluster dominated by the fusion biopsy pathway $(10,\,119)$, 26.5\% (9/34; 95\% CI 14.6--43.1\%) in the cluster dominated by the open prostatectomy pathway $(106)$, and 7.3\% (7/96; 95\% CI 3.6--14.3\%) in the cluster dominated by the robotic-assisted prostatectomy pathway $(106,\,119)$ (primary split; 95\% Wilson confidence intervals; Fisher exact test, $p<0.001$; all pairwise differences significant after Holm adjustment, $p<0.001$, $p<0.001$ and $p=0.006$). Patients whose first stay was the diagnostic fusion biopsy stay were thus almost always admitted again, consistent with surgery during a separate admission, whereas patients whose first stay already included the prostatectomy were admitted again much less often. Across the 20 splits, the association of our clusters with a subsequent admission (median Cram\'er's~$V$ 0.84) was similar to that of LDA on single codes (0.83) and higher than for all other methods (0.42--0.80), including k-means on transition frequencies (0.61); with freely learned components, it dropped to 0.54. A subsequent admission includes planned and unplanned admissions and therefore does not necessarily indicate a complication. As these clusters are based on truncated trajectories, they are not identical to those reported in Section~4.3.}

\begin{table}[ht]
\centering
\caption{Typical-pathway graph vs.\ directly-follows graph (DFG) of the same size, both derived from the training patients of each of 20 train--test splits ($\tau = 15$; median over splits). Held-out coverage: share of test transitions that are edges of the graph. Stability: edge Jaccard similarity between the graphs of different splits (190 pairs).}

\label{R1}
\begin{tabular}{lcc}
\hline
 & \textbf{Typical-pathway graph} & \textbf{DFG} \\
\hline
Number of transitions          & 16     & 16     \\
Held-out coverage              & 72.0\% & 72.7\% \\
Stability (edge Jaccard)       & 0.94   & 0.88   \\
Splits with cycles             & 0/20   & 20/20  \\
Splits with unreachable states & 0/20   & 10/20  \\
\hline
\end{tabular}
\end{table}

\begin{table}[ht]
\centering
\caption{Stability of the patient clusters across methods (complete trajectories, $\tau = 15$); median over 190 split pairs from 20 train--test splits. Adjusted Rand index (ARI) and normalized mutual information (NMI) between the clusterings of all patients obtained from different training sets.}
\label{R2}
\begin{tabular}{lcc}
\hline
\textbf{Method} & \textbf{ARI} & \textbf{NMI} \\
\hline
\textbf{Our framework} ($\tau = 15$) & \textbf{0.81} & \textbf{0.75} \\
Ablation (freely learned components) & 0.46 & 0.46 \\
Mixture of Markov chains & 0.46 & 0.40 \\
Mixture of HMMs & 0.70 & 0.59 \\
LDA, transition tokens & 0.46 & 0.46 \\
LDA, single codes & 0.31 & 0.38 \\
Edit distance, k-medoids & 0.70 & 0.77 \\
k-means, transition frequencies & 0.62 & 0.69 \\
\hline
\end{tabular}

\end{table}

\section{Discussion} \label{Discussion}

This study introduces a framework for analyzing heterogeneous healthcare trajectories by combining cohort-level pathway identification with patient-level admixture modeling and clustering. Applied to routine care data from prostate cancer patients undergoing radical prostatectomy, the framework produced stable pathway structures and interpretable patient-level trajectory representations across multiple train–test splits. A central feature of the framework is the separation of cohort-level pathway identification from patient-level trajectory modeling. Rather than assigning patients to a single pathway, the admixture model represents each trajectory as a probabilistic combination of typical pathways, allowing heterogeneous and intermediate care patterns to be captured. The resulting admixture vectors provide a compact and comparable representation of patient trajectories for visualization and clustering. \revone{5}{Empirically, the framework demonstrated robustness to sampling variability. Across multiple random train–test splits, the same typical pathway structures, comparable admixture representations, and stable cluster patterns were consistently observed, supporting the framework for exploratory analysis of routine care data}{Across 20 random train–test splits, the same three initial pathway branches were recovered in every split, the displayed pathway transitions overlapped substantially, and the patient clusters were reproducible, supporting the use of the framework for exploratory analysis of routine care data.} An important question is whether the patient subgroups identified in the admixture space are clinically meaningful. In the prostate cancer cohort, the three identified clusters reflected distinct and clinically interpretable care trajectory patterns. Cluster 1 primarily corresponded to patients undergoing \revone{3}{more extensive in-hospital diagnostic workup prior to surgery, including fusion biopsy, whereas Cluster 2 reflected a more direct trajectory to robotic-assisted prostatectomy, representing patients with externally completed staging. Cluster 3 captured patients with mixed pathway contributions and potentially more complex perioperative courses.}{in-hospital diagnostic workup, including fusion biopsy, before surgery; in the analysis of trajectories up to the first discharge, the corresponding cluster was almost always followed by a subsequent admission, consistent with surgery during a separate admission (Section~\ref{4.4}). Cluster 2 reflected a more direct trajectory to robotic-assisted prostatectomy, consistent with externally completed staging. Cluster 3 captured patients with mixed pathway contributions.}

\revone{4}{}{The comparison with alternative approaches shows where the advantages of our framework lie. Unlike existing approaches, which either describe the care process of a cohort (process mining) or group patients without an explicit pathway structure (topic models, mixture models, sequence clustering), the framework provides both in one consistent
representation. At the cohort level, the typical-pathway graph had a size and coverage
comparable to a filtered process-mining graph, but organized care into stage-specific
pathways without cycles. At the patient level, using these pathways as model components made the representation more stable and more clinically associated than freely learned components; its clusters were more stable than those of the other probabilistic models, and their association with a subsequent admission was among the highest of all approaches. Because the components are fixed to the typical pathways rather than fitted to the individual trajectories, models with freely estimated components, including our own ablation, reproduce individual trajectories more closely; the framework accepts this loss of fit in exchange for stability, interpretability, and clinical association. $k$-means on transition frequencies produced less stable partitions on this cohort and did not provide pathway-level interpretation.}
Beyond subgroup characterization, the admixture vector $q_i$ may also provide clinically relevant information at the individual patient level by summarizing care complexity not immediately visible from raw event sequences. Admixture weights may support benchmarking of patient trajectories against cohort-level norms and contribute to quality assurance and care coordination. However, translating these representations into actionable clinical decisions requires careful contextual interpretation and integration with clinical expertise, and the framework should currently be regarded as a tool for retrospective exploratory analysis rather than prospective guidance.
We further like to point out that real-world pathway representations may also provide a complementary source of evidence for understanding how care is delivered in practice. By systematically identifying frequently observed care trajectories, the framework may support comparisons between guideline-based recommendations and actual clinical practice, thereby helping to identify recurrent care patterns across institutions and patient populations. In addition, the framework may facilitate the systematic study of variation in healthcare delivery. Differences between identified pathway structures may reflect intended variation arising from patient-specific treatment decisions and personalized care, but may also highlight unintended variation associated with inconsistent practice patterns. Consequently, pathway-based representations derived from routine care data may become increasingly relevant for quality assurance and quality improvement initiatives. An additional strength of the proposed framework is that it relies on routinely collected administrative and clinical coding information, including diagnoses, procedures, and timestamps. Such data are widely available in many healthcare systems and, for example, are routinely collected in Germany through the §21 hospital billing dataset, potentially facilitating broader application of the framework across institutions and clinical settings. Still, several steps would be required before deployment in routine clinical settings. \revtwo{1}{While the framework produced stable results in the present cohort, its generalizability across institutions, patient populations, and coding practices remains to be evaluated, as routine care data may vary substantially across centers.}{} Practical implementation would also require integration with electronic health record systems and robust handling of incomplete or inconsistently recorded data.

Certain limitations should be considered when interpreting the findings. \revtwo{1}{}{The clinical evaluation was restricted to patients undergoing radical prostatectomy in the Department of Urology at the University Medical Center Freiburg. The framework itself does not depend on a particular coding system: it uses coded events only as state labels and derives pathways and patient representations from the order and frequency of events, without code-specific rules; clinical knowledge is needed only to interpret the resulting pathways. It can therefore, in principle, be applied to other cohorts recorded as sequences of coded events. The clinical findings, however, i.e.\ the identified pathways and subgroups, reflect the care organization, referral patterns and coding practice of a single department, and their generalizability to other institutions, specialties and coding systems remains untested. Differences in patient populations, care organization, coding conventions and documentation practices could affect both the identified pathway structures and the resulting admixture representations. The random train--test splits assess sampling stability within the observed cohort, while the simulation study evaluates performance under specified synthetic conditions; neither establishes external clinical validity. Application in other settings requires explicit mapping of local codes and event definitions, assessment of preprocessing and simplification settings, and validation of the resulting pathway representations and subgroup interpretations.} \revtwo{5}{}{Moreover, the identified pathways should be interpreted as recurrent structures in observed coded healthcare trajectories rather than as direct representations of latent clinical states. Routine administrative data are influenced not only by clinical care but also by coding, billing, and documentation practices. In particular, recurring code bundles and identical timestamps may reflect genuine clinical co-occurrence, documentation conventions, or a combination of both. Similarly, discharge is an administrative event that defines the observed end of an inpatient trajectory. Consequently, some differences between the identified pathway structures may reflect institution-specific documentation practices in addition to differences in actual care delivery, and clinical interpretations of individual pathway states and patient subgroups should be made with this distinction in mind.} The cohort-level pathway algorithm is descriptive and depends on collapsing and pruning thresholds that influence the structural resolution of the pathway graph, although in our application these parameters mainly affected granularity rather than the identity of the typical pathway structures. \revone{5}{}{The stability analysis assesses within-cohort sampling variation at the level of displayed simplified pathway transitions. It does not establish identical complete graph topology or stability of all event-block labels stored within merged nodes.} \revtwo{4}{}{Because the repeated random train--test splits were drawn from the same center and observation period, they assess sampling stability within the observed cohort, but not robustness to temporal, geographic, or workflow shifts. Assessing such robustness would require separate validation designs, such as training on earlier observations and evaluating on a later period, evaluating an independent institutional cohort, or examining performance before and after a documented change in care delivery or coding practice.} Furthermore, the admixture model is based on first-order Markov chains derived from the simplified pathway graph. While this formulation captures dominant local transition dynamics and remains tractable in heterogeneous routine care data, it does not model longer-range temporal dependencies or context-specific effects spanning multiple care steps. Although deep learning approaches may capture richer temporal dependencies [4], first-order Markov and mixed-membership models provide an interpretable and stable representation of routine care trajectories. \revone{4}{}{In the comparison with alternative approaches, all methods were fitted with $K = 3$, the number of components of our framework, which need not be optimal for every method; for mixtures of hidden Markov models, the Bayesian information criterion selected the largest number of hidden states considered, so that larger models might have performed differently.} A further limitation is that the analysis is retrospective and based on completed hospital trajectories leading to discharge. Consequently, the framework characterizes pathway structure and patient-level trajectory composition after the full care process has been observed rather than supporting real-time clinical decision-making. Extending the approach to prospective settings would require additional methodological development and validation beyond the scope of the present study.

\section{Conclusion}
In this work, we presented a framework for analyzing heterogeneous healthcare trajectories by combining cohort-level typical pathway identification with patient-level admixture modeling. Applied to routine care data from prostate cancer patients undergoing radical prostatectomy, the framework produced a stable and interpretable typical healthcare pathway and enabled the identification of patient subgroups with distinct care progression patterns. By representing patient trajectories as probabilistic mixtures of typical pathways, the approach captures both shared and intermediate treatment patterns while reducing complexity in routine care data. The proposed framework is particularly applicable to the retrospective analysis of heterogeneous real-world healthcare trajectories, where it enables the identification of typical care patterns and clinically interpretable patient subgroups.

\section*{CRediT authorship contribution statement}
\noindent
\textbf{Maryam Farhadizadeh}: Conceptualization, Methodology, Software, Formal analysis, Investigation, Data Curation, Writing -- original draft, Visualization. 
\textbf{Carola S. Heinzel}: Methodology, Software, Writing -- review \& editing. 
\textbf{August Sigle}: Writing -- review \& editing. 
\textbf{Harald Binder}: Methodology, Writing -- review \& editing. 
\textbf{Frederik Wenz}: Writing -- review \& editing. 
\textbf{Jan Hasenauer}: Writing -- review \& editing.
\textbf{Peter Pfaffelhuber}: Methodology, Writing -- review \& editing. 
\textbf{Nadine Binder}: Conceptualization, Supervision, Methodology, Writing -- review \& editing, Funding acquisition.

\section*{Funding sources}
\noindent
This work was funded by the Deutsche Forschungsgemeinschaft (DFG, German Research Foundation) — Project-ID 499552394 — SFB 1597, and the European Union via the ERC grant INTEGRATE — Project-ID 101126146. The funding source had no involvement in the study design; collection, analysis, or interpretation of data; writing of the manuscript; or the decision to submit the article for publication.

\section*{Declaration of competing interest}
\noindent
The authors declare that they have no known competing financial interests or personal relationships that could have appeared to influence the work reported in this paper.

\section*{Declaration of Generative AI and AI-assisted technologies in the writing process}
\noindent
During the preparation of this work, the authors used generative AI tools to improve language, readability, and text organization. After using these tools, the authors reviewed and edited the content as needed and take full responsibility for the content of the published article.

\section*{Ethics statement}
\noindent
The protocol for the analysis of the data in this work was approved by the Ethics Committee, Medical Center, University of Freiburg, Freiburg, Germany (reference number 22-1433-S1-retro). The committee waived the requirement for informed consent because the data were analyzed anonymously.

\section*{Data availability}
\noindent
The clinical routine care data analyzed in this study are not publicly available due to data protection and institutional regulations.

\section*{Code availability}
\noindent
\revtwo{6}{The current version of the code, available on GitHub, is implemented for simulated data and can be adapted for other datasets:
\url{https://github.com/MaryamFarhadizadeh/HealthcarePathway-AdmixtureModel}}{The simulation code and reproducibility materials are publicly available at \url{https://github.com/MaryamFarhadizadeh/HealthcarePathway-AdmixtureModel}. The repository provides a documented, executable workflow covering synthetic data generation, pathway discovery, admixture estimation, clustering, and recovery evaluation. It includes pinned software dependencies, complete scenario configurations, random seeds, and execution instructions. The workflow requires no access to the clinical records. Clinical data and private real-data analysis materials cannot be made publicly available because of data-protection restrictions.}

\section*{Acknowledgments}
\noindent
Some icons used in Figure \ref{fig:workflow} were designed by authors from Flaticon (\url{https://www.flaticon.com/}).

\section*{Supplementary data}
\noindent
Supplementary material related to this article can be found online
at (weblink to be added later).

%%%%%%%%%%%%%%%%%%%%%%%%%%%%%%%%%%%%%%%%%%%%%%%%%%%
% REFERENCES

\bibliographystyle{elsarticle-num}
\bibliography{JBI_lit}

@article{aspland2021clinical,
  title   = {Clinical pathway modelling: a literature review},
  author  = {Aspland, Emma and Gartner, Daniel and Harper, Paul},
  journal = {Health Systems},
  volume  = {10},
  number  = {1},
  pages   = {1--23},
  year    = {2021}
}

@article{huang2018probabilistic,
  title   = {Probabilistic modeling personalized treatment pathways using electronic health records},
  author  = {Huang, Zhengxing and Ge, Zhenxiao and Dong, Wei and He, Kunlun and Duan, Huilong},
  journal = {Journal of Biomedical Informatics},
  volume  = {86},
  pages   = {33--48},
  year    = {2018}
}

@article{perer2015mining,
  title   = {Mining and exploring care pathways from electronic medical records with visual analytics},
  author  = {Perer, Adam and Wang, Fei and Hu, Jianying},
  journal = {Journal of Biomedical Informatics},
  volume  = {56},
  pages   = {369--378},
  year    = {2015},
  doi     = {10.1016/j.jbi.2015.06.020}
}

@article{rosa2022modelling,
  title   = {Modelling and mining of patient pathways: A scoping review},
  author  = {Rosa, Caroline de Oliveira Costa Souza and Ito, Márcia and Vieira, Alex Borges and Gomes, Antônio Tadeu Azevedo},
  journal = {arXiv},
  eprint  = {2206.01980},
  archivePrefix = {arXiv},
  primaryClass = {cs.LG},
  year    = {2022}
}

@article{li2024temporal,
  title   = {Temporal topic model for clinical pathway mining from electronic medical records},
  author  = {Li, Wei and Min, Xin and Ye, Panpan and Xie, Weidong and Zhao, Dazhe},
  journal = {BMC Medical Informatics and Decision Making},
  volume  = {24},
  number  = {1},
  pages   = {20},
  year    = {2024},
  doi     = {10.1186/s12911-024-02384-3}
}

@inproceedings{xu2016tcpm,
  title     = {{TCPM}: Topic-Based Clinical Pathway Mining},
  author    = {Xu, Xiao and Jin, Tao and Wei, Zhijie and Lv, Cheng and Wang, Jianmin},
  booktitle = {Proceedings of the 2016 IEEE First International Conference on Connected Health: Applications, Systems and Engineering Technologies (CHASE)},
  pages     = {292--301},
  year      = {2016},
  publisher = {IEEE},
  doi       = {10.1109/CHASE.2016.43}
}

@article{muyama2024machine,
  title={Machine learning approaches for the discovery of clinical pathways from patient data: A systematic review},
  author={Muyama, Lilia and Neuraz, Antoine and Coulet, Adrien},
  journal={Journal of Biomedical Informatics},
  volume={145},
  pages={104451},
  year={2024},
  publisher={Elsevier},
  doi={10.1016/j.jbi.2024.104451}
}

@article{najjar2018two,
  title        = {A two-step approach for mining patient treatment pathways in administrative healthcare databases},
  author       = {Najjar, Amine and Reinharz, Daniel and Girouard, Claude and Gagn{\'e}, Christophe},
  journal      = {Artificial Intelligence in Medicine},
  volume       = {87},
  pages        = {34--48},
  year         = {2018},
  doi          = {10.1016/j.artmed.2018.03.004}
}

@article{xiao2018opportunities,
  title={Opportunities and challenges in developing deep learning models using electronic health records data: a systematic review},
  author={Xiao, Cao and Choi, Edward and Sun, Jimeng},
  journal={Journal of the American Medical Informatics Association},
  volume={25},
  number={10},
  pages={1419--1428},
  year={2018},
  publisher={Oxford University Press},
  doi={10.1093/jamia/ocy068}
}

@article{binder2022,
  title={Data mining in urology: Understanding real-world treatment pathways for lower urinary tract systems via exploration of big data},
  author={Binder, Nadine and Dette, Holger and Franz, Julia and Z{\"o}ller, Daniela and Suarez-Ibarrola, Rodrigo and Gratzke, Christian and Binder, Harald and Miernik, Arkadiusz},
  journal={European Urology Focus},
  volume={8},
  number={2},
  pages={391--393},
  year={2022},
  publisher={Elsevier}
}

@article{rabiner1989tutorial,
  title={A tutorial on hidden Markov models and selected applications in speech recognition},
  author={Rabiner, Lawrence R.},
  journal={Proceedings of the IEEE},
  volume={77},
  number={2},
  pages={257--286},
  year={1989},
  publisher={IEEE}
}

@inproceedings{macqueen1967some,
  title={Some methods for classification and analysis of multivariate observations},
  author={MacQueen, J.},
  booktitle={Proceedings of the Fifth Berkeley Symposium on Mathematical Statistics and Probability},
  volume={1},
  pages={281--297},
  year={1967},
  publisher={University of California Press}
}

@article{blei2003latent,
  title={Latent Dirichlet Allocation},
  author={Blei, David M. and Ng, Andrew Y. and Jordan, Michael I.},
  journal={Journal of Machine Learning Research},
  volume={3},
  pages={993--1022},
  year={2003}
}

@article{rismanchian2023data,
  title={A data-driven approach to support the understanding and improvement of patients’ journeys: a case study using electronic health records of an emergency department},
  author={Rismanchian, Farzad and Kassani, S. H. and Shavarani, S. M. and Lee, Young Hae},
  journal={Value in Health},
  volume={26},
  number={1},
  pages={18--27},
  year={2023},
  publisher={Elsevier}
}

@inproceedings{quintano2019sepsis,
  title={Analysis and optimization of a sepsis clinical pathway using process mining},
  author={Quintano Neira, Rodrigo A. and Hompes, Bart F.A. and De Vries, Jeroen G.J. and Mazza, Bruna F. and Sim{\~o}es de Almeida, Samuel L. and Stretton, Emma and Buijs, Joos C. and Hamacher, Silvio},
  booktitle={International Conference on Business Process Management},
  pages={459--470},
  year={2019},
  publisher={Springer International Publishing},
  address={Cham}
}

@article{chiudinelli2020mining,
  title={Mining post-surgical care processes in breast cancer patients},
  author={Chiudinelli, Lorenzo and Dagliati, Arianna and Tibollo, Veronica and Albasini, Silvia and Geifman, Nophar and Peek, Niels and Holmes, John H. and Corsi, Fabio and Bellazzi, Riccardo and Sacchi, Lucia},
  journal={Artificial Intelligence in Medicine},
  volume={105},
  pages={101855},
  year={2020},
  publisher={Elsevier}
}

@article{huang2019medication,
  title={Discovering medication patterns for high-complexity drug-using diseases through electronic medical records},
  author={Huang, Hong and Shang, Xiaofeng and Zhao, Haibin and Wu, Ning and Li, Wei and Xu, Yizhou and Zhou, Ying and Fu, Lei},
  journal={IEEE Access},
  volume={7},
  pages={125280--125299},
  year={2019},
  publisher={IEEE}
}

@inproceedings{lin2021personalized,
  title={Personalized clinical pathway recommendation via attention based pre-training},
  author={Lin, Xiaobo and Li, Yifan and Xu, Yizhou and Guo, Wei and He, Wenjun and Zhang, Huan and Cui, Lei and Miao, Chunyan},
  booktitle={2021 IEEE International Conference on Bioinformatics and Biomedicine (BIBM)},
  pages={980--987},
  year={2021},
  organization={IEEE}
}

@article{ye2020predicting,
  title={Predicting optimal hypertension treatment pathways using recurrent neural networks},
  author={Ye, Xi and Zeng, Qing T. and Facelli, Julio C. and Brixner, Diana I. and Conway, Mike and Bray, Bruce E.},
  journal={International Journal of Medical Informatics},
  volume={139},
  pages={104122},
  year={2020},
  publisher={Elsevier}
}

@article{muyama2024drl,
  title={Deep reinforcement learning for personalized diagnostic decision pathways using electronic health records: A comparative study on anemia and systemic lupus erythematosus},
  author={Muyama, L. and Neuraz, A. and Coulet, A.},
  journal={Artificial Intelligence in Medicine},
  volume={157},
  pages={102994},
  year={2024},
  publisher={Elsevier}
}

@book{vanderAalst2011,
  title={Process Mining: Discovery, Conformance and Enhancement of Business Processes},
  author={van der Aalst, Wil M. P.},
  year={2011},
  publisher={Springer},
  address={Berlin, Heidelberg}
}

@article{levenshtein1966binary,
  title={Binary codes capable of correcting deletions, insertions and reversals},
  author={Levenshtein, Vladimir I.},
  journal={Soviet Physics Doklady},
  volume={10},
  number={8},
  pages={707--710},
  year={1966}
}

@book{norris1998markov,
  title={Markov Chains},
  author={Norris, J. R.},
  year={1998},
  publisher={Cambridge University Press},
  address={Cambridge}
}

@Article{alexander2009fast,
  author    = {Alexander, David H and Novembre, John and Lange, Kenneth},
  journal   = {Genome research},
  title     = {Fast model-based estimation of ancestry in unrelated individuals},
  year      = {2009},
  number    = {9},
  pages     = {1655--1664},
  volume    = {19},
  comment   = {ADMIXTURE, Anwendung: Population Stratisfication},
  publisher = {Cold Spring Harbor Lab},
}

@Article{pritchard2000,
  author    = {Pritchard, Jonathan K and Stephens, Matthew and Donnelly, Peter},
  journal   = {Genetics},
  title     = {Inference of population structure using multilocus genotype data},
  year      = {2000},
  number    = {2},
  pages     = {945--959},
  volume    = {155},
  comment   = {STRUCTURE},
  publisher = {Oxford University Press},
}

@article{funkner_towards_2017,
	title = {Towards evolutionary discovery of typical clinical pathways in electronic health records},
	volume = {119},
	issn = {1877-0509},
	journal = {Procedia computer science},
	author = {Funkner, Anastasia A and Yakovlev, Aleksey N and Kovalchuk, Sergey V},
	year = {2017},
	note = {Publisher: Elsevier},
	pages = {234--244},
}

@article{lawal2016clinical,
  title={What is a clinical pathway? Refinement of an operational definition to identify clinical pathway studies for a Cochrane systematic review},
  author={Lawal, A. K. and Rotter, T. and Kinsman, L. and Machotta, A. and Ronellenfitsch, U. and Scott, S. D. and Goodridge, D. and Plishka, C. and Groot, G.},
  journal={BMC Medicine},
  volume={14},
  number={1},
  pages={35},
  year={2016},
  publisher={BioMed Central},
  doi={10.1186/s12916-016-0580-z}
}

@article{semler2018mii,
  title     = {The German Medical Informatics Initiative},
  author    = {Semler, S. C. and Wissing, F. and Heyder, R.},
  journal   = {Methods of Information in Medicine},
  volume    = {57},
  number    = {S 01},
  pages     = {e50--e56},
  year      = {2018},
  publisher = {Schattauer},
  doi       = {10.3414/ME18-03-0003}
}

@article{evanno2005,
  title     = {Detecting the Number of Clusters of Individuals Using the Software STRUCTURE: A Simulation Study},
  author    = {Evanno, G. and Regnaut, S. and Goudet, J.},
  journal   = {Molecular Ecology},
  volume    = {14},
  number    = {8},
  pages     = {2611--2620},
  year      = {2005},
  doi       = {10.1111/j.1365-294X.2005.02553.x}
}

@article{alexander2009,
  title     = {Fast Model-Based Estimation of Ancestry in Unrelated Individuals},
  author    = {Alexander, D. H. and Novembre, J. and Lange, K.},
  journal   = {Genome Research},
  volume    = {19},
  number    = {9},
  pages     = {1655--1664},
  year      = {2009},
  doi       = {10.1101/gr.094052.109}
}
\section*{Supplementary Material}

\renewcommand{\thefigure}{S\arabic{figure}}
\setcounter{figure}{0}

\renewcommand{\thetable}{S\arabic{table}}
\setcounter{table}{0}

\renewcommand{\thesection}{S\arabic{section}}

\section{Simulation study}\label{sec:supp_simulation}
% =====================================================================================

To evaluate the ability of the proposed pathway admixture framework to recover latent pathway structure and patient-level mixture proportions under controlled conditions, we conducted a simulation study with progressively increasing structural complexity. All simulations were conducted with $N=400$ patients.

\subsection{Simulation design}\label{sec:supp_sim_design}

We defined three latent backbone pathways, each represented as an ordered sequence of states from a common start state $a$ to a common end state $e$:
\begin{align*}
B_1 &: a \rightarrow b \rightarrow f \rightarrow h \rightarrow e, \\
B_2 &: a \rightarrow c \rightarrow g \rightarrow h \rightarrow e, \\
B_3 &: a \rightarrow d \rightarrow i \rightarrow h \rightarrow e.
\end{align*}

The backbones share the start and end states as well as one intermediate state, while differing in internal transitions. This partial structural overlap ensures that pathway discrimination requires modeling transition dynamics rather than relying on trivial state separation.

For each simulated patient $i$, a latent mixture vector
\[
\boldsymbol{\theta}_i = (\theta_{i1}, \theta_{i2}, \theta_{i3})
\]
was sampled from a symmetric Dirichlet distribution,
\[
\boldsymbol{\theta}_i \sim \text{Dirichlet}(\alpha),
\]
where the concentration parameter $\alpha$ controls sparsity of mixture weights. Smaller values of $\alpha$ yield near-pure pathway membership, whereas larger values produce more diffuse mixtures.

The dominant backbone label was defined as
\[
z_i = \arg\max_k \theta_{ik},
\]
and was used for discrete recovery evaluation.

Patient trajectories were constructed position-wise along the backbone length. At each internal position, the generating backbone source was selected as the dominant backbone $z_i$ with probability $1 - p_{\text{switch}}$, or sampled according to the mixture weights $\boldsymbol{\theta}_i$ with probability $p_{\text{switch}}$. This mechanism induces controlled within-sequence switching across backbone sources, generating trajectories that represent finite mixtures of pathway-specific transition processes. The switching probability $p_{\text{switch}}$ therefore directly governs structural ambiguity.

To emulate real-world clinical variability, we additionally introduced stochastic perturbations including node skipping, node repetition, insertion of auxiliary codes within event blocks (vertical noise), and insertion of transient states between transitions (horizontal noise). A backbone-specific anchor code was further inserted with fixed probability, providing a weak global signal linked to the dominant backbone.

These perturbations were chosen to reflect typical sources of variability observed in routine healthcare data, including concurrent coding, documentation timing differences, and minor deviations in care progression. This design enables evaluation of structural recovery under realistic levels of heterogeneity rather than idealized, noise-free conditions.

\begin{figure}[htbp]
\centering
\begin{subfigure}[b]{0.48\textwidth}
    \centering
    \includegraphics[width=\linewidth]{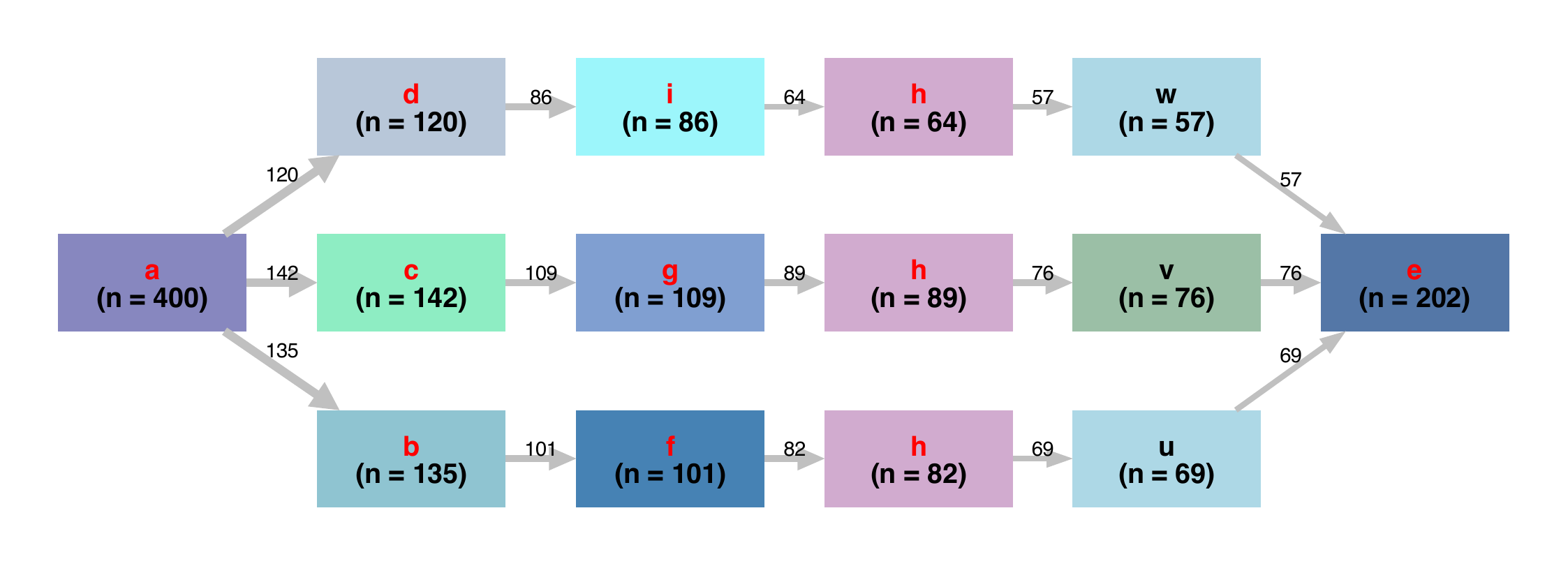}
    \caption{Low structural mixing}
\end{subfigure}
\hfill
\begin{subfigure}[b]{0.50\textwidth}
    \centering
    \includegraphics[width=\linewidth]{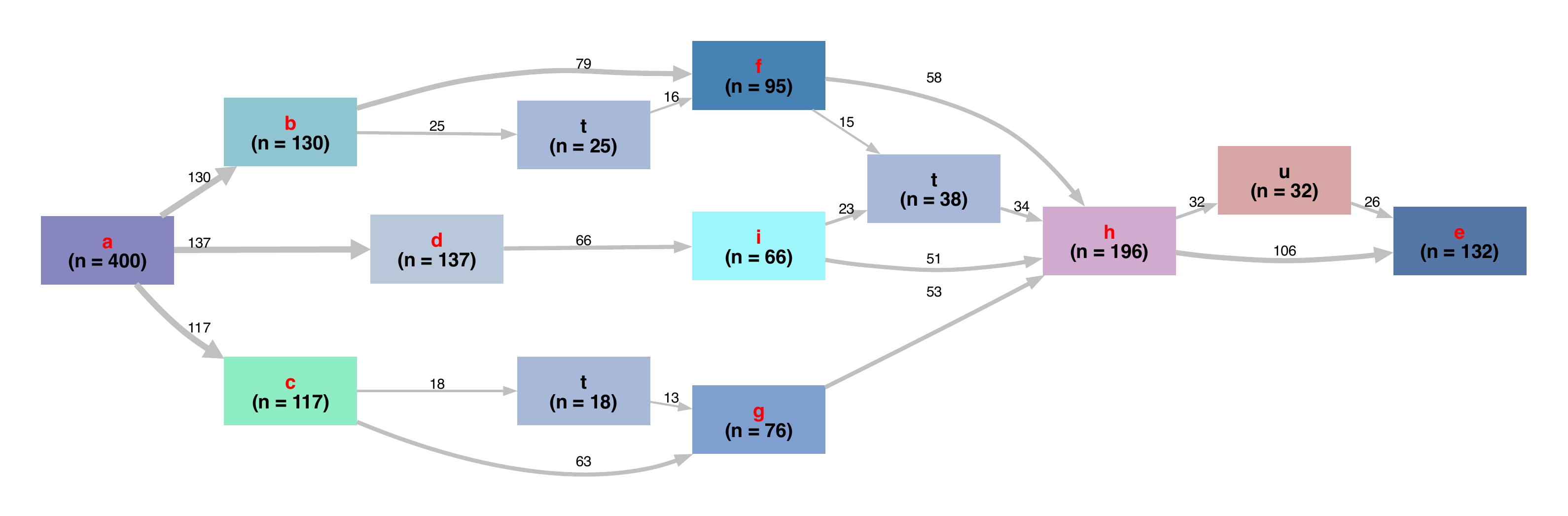}
    \caption{Moderate structural mixing}
\end{subfigure}

\vspace{0.4cm}
\begin{subfigure}[b]{0.75\textwidth}
    \centering
    \includegraphics[width=\linewidth]{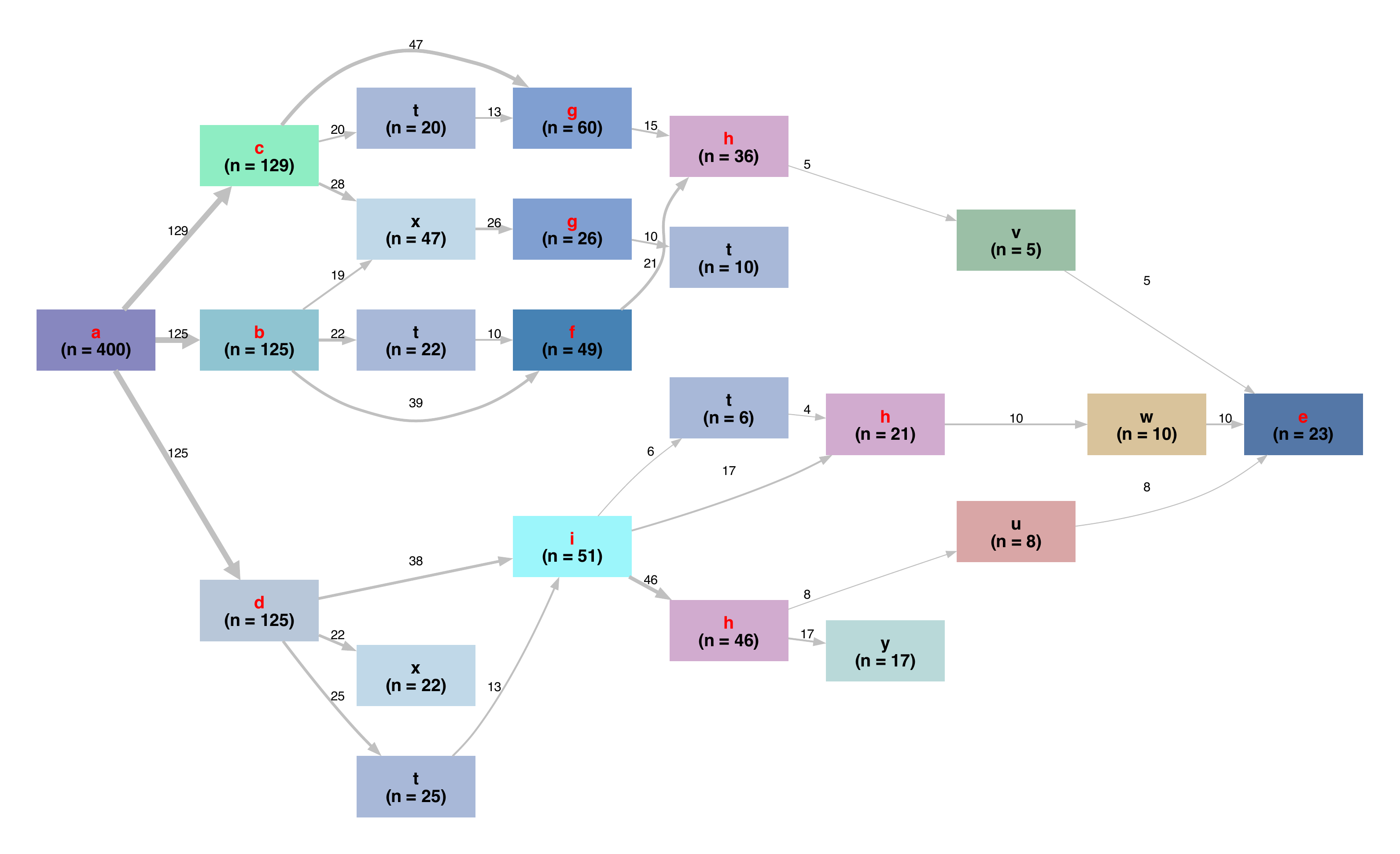}
    \caption{High structural mixing}
\end{subfigure}
\caption{Structural recovery of latent backbone pathways under increasing mixing intensity.
The simplified prefix tree extracted from simulated data reveals three principal branches corresponding to the underlying generative backbones.
As switching probability and mixture diffuseness increase from (a) low to (c) high structural mixing, additional transient nodes and branch overlap emerge; nevertheless, the primary backbone topology remains recoverable.}
\label{fig:simulation_structure}
\end{figure}

Consistent with the generative design, the simplified pathway structures reveal three principal branches corresponding to the latent backbones across all structural mixing regimes (Figure~\ref{fig:simulation_structure}). Under low structural mixing, the recovered branches align directly with the underlying generative backbones. As switching intensity increases, additional transient nodes and cross-branch transitions emerge, resulting in increased structural overlap. Nevertheless, even under high structural mixing, the primary backbone topology remains identifiable, indicating that the simplification procedure preserves core pathway structure despite substantial within-sequence switching.

Three levels of increasing complexity were defined by jointly varying the Dirichlet concentration parameter and the switching probability (Table~\ref{tab:simulation_parameters}). Increasing $\alpha$ results in more diffuse mixture weights, while increasing $p_{\text{switch}}$ increases within-trajectory structural mixing. The probabilities of the stochastic perturbations also increase from the low to the high regime; the complete configurations of all three regimes are provided with the simulation code.

\begin{table}[htbp]
\centering
\caption{Simulation regimes with increasing structural mixing induced by higher Dirichlet concentration and switching probability.}
\label{tab:simulation_parameters}
\begin{tabular}{lcc}
\toprule
Structural mixing level & Dirichlet $\alpha$ & Switching probability $p_{\text{switch}}$ \\
\midrule
Low       & 0.8 & 0.10 \\
Moderate  & 1.0 & 0.22 \\
High      & 1.2 & 0.35 \\
\bottomrule
\end{tabular}
\end{table}

% Reviewer 2, Comment 6
All simulation experiments were performed using the documented reproducibility workflow. Data generation used random seed 42, patient-level train--test partitions used seeds 35, 123, and 2025, and $k$-means clustering used random seed 42. For each simulation scenario, one synthetic cohort was generated and the three train--test partitions were subsequently applied to that cohort. Complete scenario configurations, software dependencies, and execution instructions are provided with the accompanying simulation code.

\subsection{Evaluation metrics}\label{sec:supp_sim_metrics}

Performance was evaluated at three complementary levels. Continuous mixture recovery was assessed by comparing estimated admixture weights $\mathbf{q}_i$ to ground-truth weights $\boldsymbol{\theta}_i$ using mean absolute error (MAE), root mean squared error (RMSE), and Pearson correlation. To ensure label invariance, component matching was performed via permutation search.

Discrete backbone recovery was evaluated by comparing the dominant estimated component $\arg\max_k q_{ik}$ with the true dominant backbone $z_i$ using the Adjusted Rand Index (ARI).

Finally, clustering quality in admixture space was assessed using the silhouette score, Calinski--Harabasz index, and Davies--Bouldin index.

We expected recovery performance and cluster separation to decrease with increasing structural mixing.

\subsection{Simulation results}\label{sec:supp_sim_results}

All reported metrics are averages across the three train--test splits (seeds 35, 123 and 2025).
Table~\ref{tab:simulation_recovery} summarizes recovery performance across the three structural mixing regimes using the EM estimator. Results for SLSQP were nearly identical (differences $<0.02$ in MAE) and are therefore omitted for brevity.

\begin{table}[htbp]
\centering
\caption{Continuous and discrete recovery performance across increasing structural mixing. MAE denotes mean absolute error between estimated admixture weights $\mathbf{q}_i$ and ground-truth weights $\boldsymbol{\theta}_i$. Correlation denotes Pearson correlation between $\mathbf{q}_i$ and $\boldsymbol{\theta}_i$. ARI denotes Adjusted Rand Index for dominant backbone recovery. Silhouette denotes cluster separation in admixture space.}
\label{tab:simulation_recovery}
\begin{tabular}{lcccc}
\toprule
Structural mixing level & MAE & Correlation & ARI & Silhouette \\
\midrule
Low       & 0.234 & 0.834 & 0.940 & 0.923 \\
Moderate  & 0.257 & 0.785 & 0.774 & 0.897 \\
High      & 0.254 & 0.784 & 0.774 & 0.876 \\
\bottomrule
\end{tabular}
\end{table}

Across all scenarios, the framework recovered the patient-level mixture weights with moderate to high accuracy. In the low structural mixing regime, the mean absolute error (MAE) was 0.234 and the correlation 0.834. From low to moderate mixing, the correlation decreased to 0.785 and the MAE increased to 0.257; from moderate to high mixing, recovery did not deteriorate further (MAE 0.254, correlation 0.784).

Recovery of the dominant backbone followed the same pattern: the Adjusted Rand Index (ARI) decreased from 0.940 under low to 0.774 under moderate mixing and remained at 0.774 under high mixing. Cluster separation in admixture space decreased across all three regimes (silhouette 0.923, 0.897 and 0.876).

A possible explanation for the plateau between the moderate and the high regime is that the higher Dirichlet concentration produces more diffuse true mixture weights, which limits the absolute estimation error. Performance did not collapse in any scenario and remained clearly below perfect recovery, so the simulation presents a non-trivial recovery task.

\FloatBarrier

% =====================================================================================
\section{Additional robustness analyses}\label{sec:supp_robustness}
% =====================================================================================

% Reviewer 1, Comment 5
\subsection{Stability across training splits}\label{sec:stability_training_splits}

To assess the stability of the framework with respect to sampling variation, we repeated the full analysis on the training patients of 20 patient-level train--test splits (80/20; random seeds 35, 123, 2025 and 1001--1017). Pathway construction and simplification were performed with $\tau = 15$ (main analysis) and, as a sensitivity analysis, with $\tau = 10$, using $\alpha = 5$ and $\beta = 0.1$ in both cases.

\paragraph{Stability of the pathway graphs}
Each simplified pathway graph was represented by its displayed nodes, identified by the procedure label shown in the figure, and its displayed directed edges. Patient counts, edge weights and event-block labels merged into a node were not used. The starting node was distinguished from later nodes with the same label. For each pair of splits (190 pairs), we computed the Jaccard similarity of the two sets of directed edges, i.e., the number of edges present in both graphs divided by the number of distinct edges present in either graph. In an additional branch-context comparison, each edge was also labelled with the initial branch from which it is reachable, so that identical transitions in different branches were counted separately. Because the 190 pairs overlap, results are summarized descriptively.

\begin{figure}[htbp]
    \centering
    \includegraphics[width=0.6\textwidth]{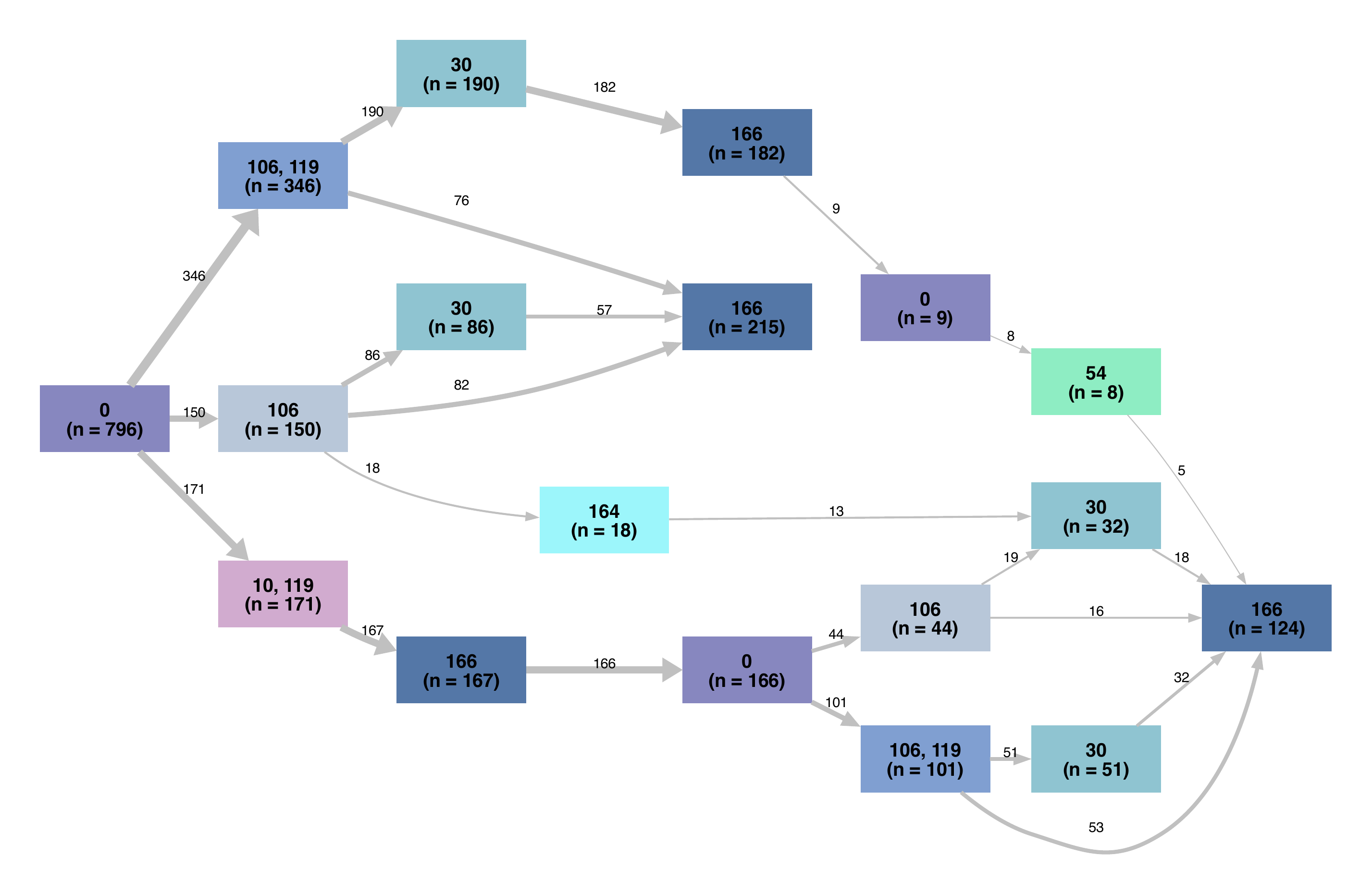}\\[0.2cm]
    \includegraphics[width=0.6\textwidth]{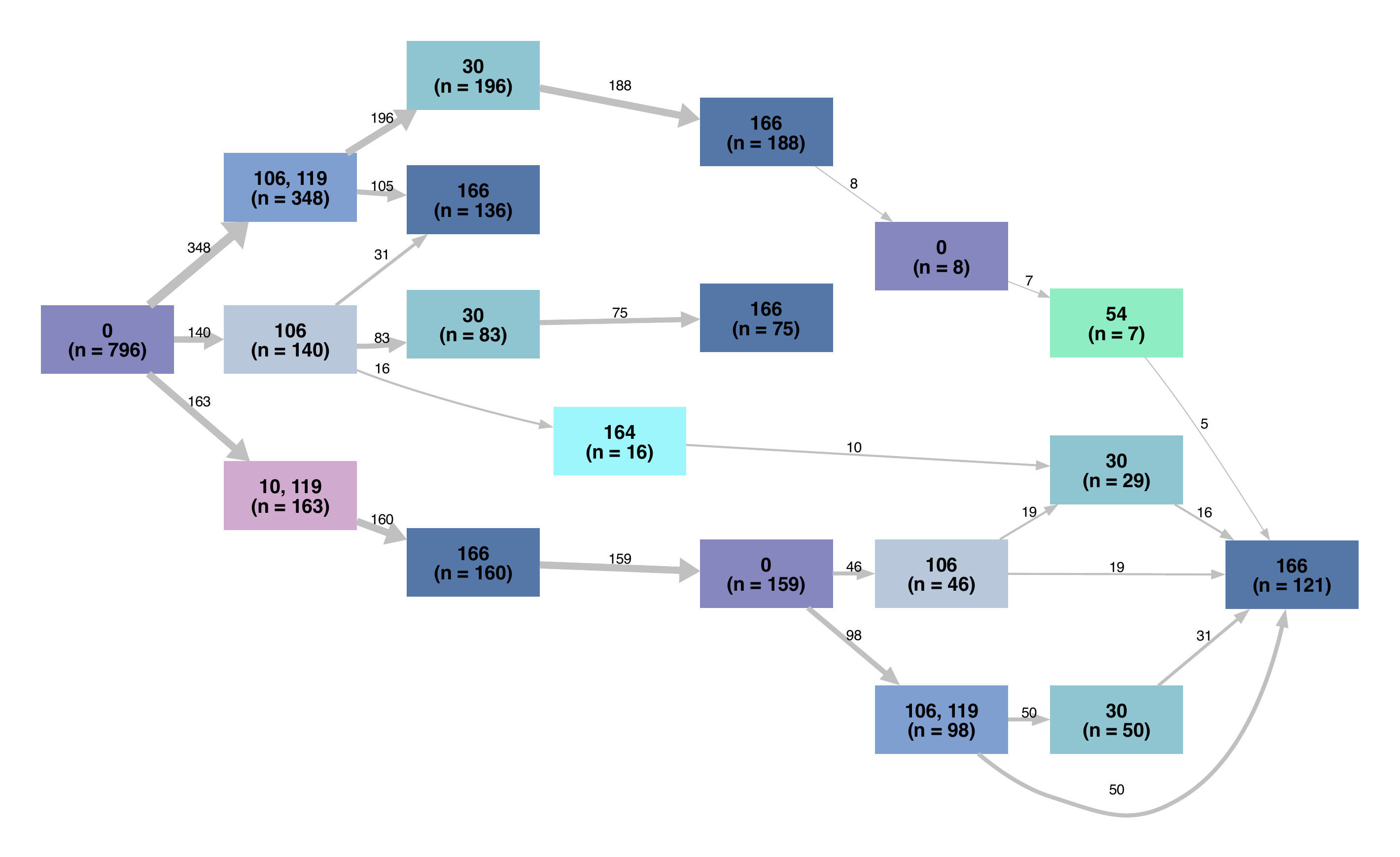}\\[0.2cm]
    \includegraphics[width=0.6\textwidth]{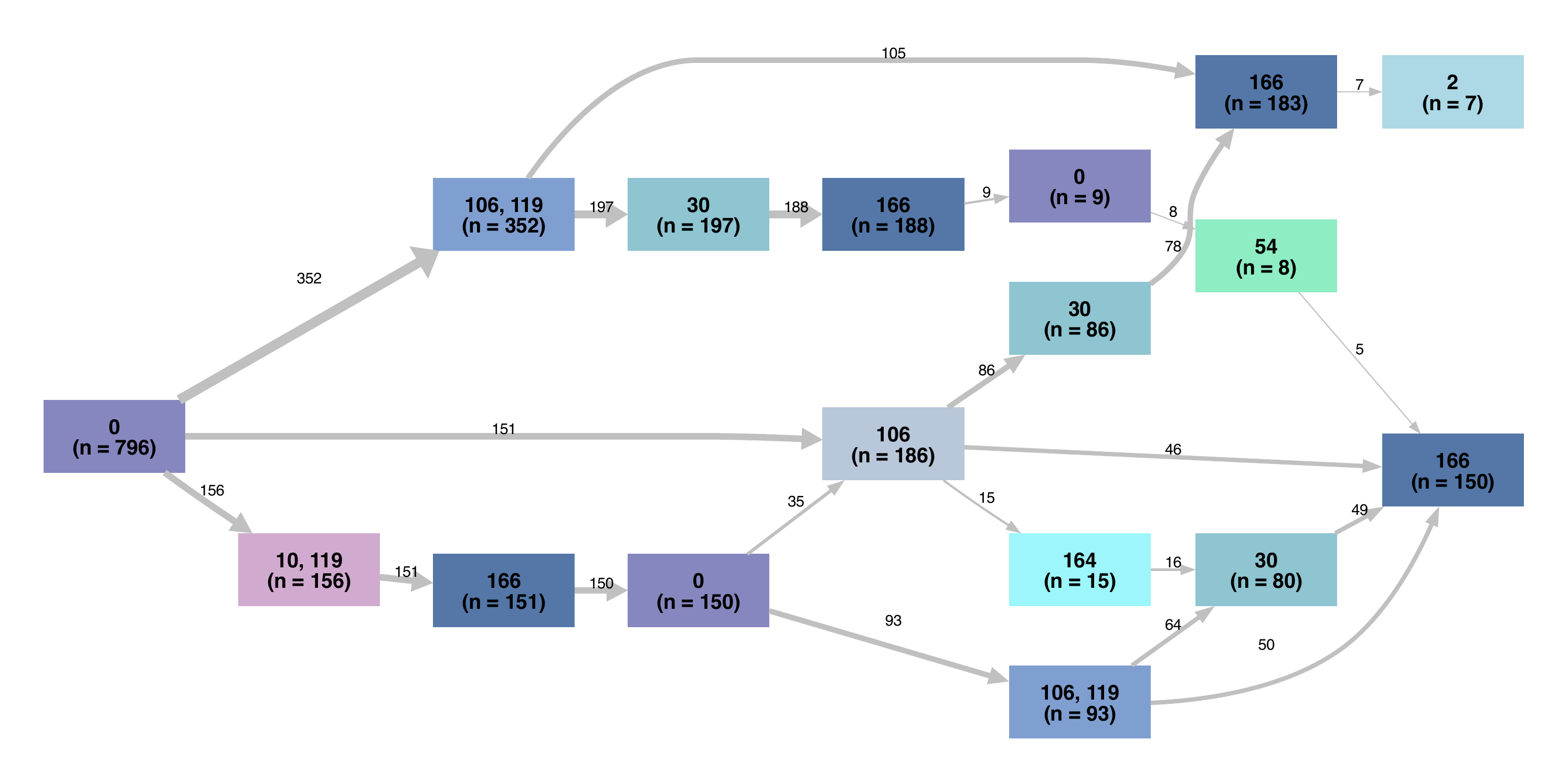}
    \caption{Examples of simplified pathway graphs derived from the training data of three of the 20 train--test splits (seeds 35, 2025 and 123). The three initial branches are the same in all three graphs; differences occur in low-support downstream branches. The quantitative comparison across all 20 splits is given in Table~\ref{tab:graph_similarity} and Figure~\ref{fig:graph_similarity}.}

    \label{fig:supp_graphs}
\end{figure}

The three initial branches, starting with $(10, 119)$, $(106)$ and $(106, 119)$, were recovered in all 20 splits for both thresholds. The pairwise similarities are summarized in Table~\ref{tab:graph_similarity} and shown in Figure~\ref{fig:graph_similarity}. The high similarity of the displayed edges indicates that the main pathway structure was reproduced consistently, while the lower minimum values reflect variation in less frequent downstream branches. Figure~\ref{fig:supp_graphs} shows the simplified pathway graphs of three of the splits as examples.

\begin{table}[htbp]
\centering
\caption{Pairwise directed-edge Jaccard similarity of the simplified pathway graphs across 20 train--test splits (190 pairs).}
\label{tab:graph_similarity}
\begin{tabular}{llccc}
\toprule
Threshold & Comparison & Median & IQR & Minimum \\
\midrule
$\tau = 15$ & Displayed edges      & 0.882 & 0.833--0.944 & 0.684 \\
$\tau = 15$ & Branch-context edges & 0.864 & 0.778--0.917 & 0.556 \\
$\tau = 10$ & Displayed edges      & 0.938 & 0.800--1.000 & 0.600 \\
$\tau = 10$ & Branch-context edges & 0.806 & 0.742--0.920 & 0.545 \\
\bottomrule
\end{tabular}
\end{table}

\paragraph{Stability of the patient clusters}
To assess the stability of the patient clusters, the pathway components, admixture weights (SLSQP) and cluster centers were estimated from the training patients of each split. All 995 patients were then assigned to the nearest cluster center, and the adjusted Rand index was computed between the resulting partitions for every pair of splits (190 pairs). The median adjusted Rand index was 0.81 (IQR 0.73--0.84) at $\tau = 15$ and 0.67 (IQR 0.59--0.75) at $\tau = 10$. The internal validation indices in Table~1 of the main text are based on the three original train--test splits.

\subsection{Sensitivity to threshold parameters}\label{sec:supp_sensitivity}

To examine the effect of the simplification parameters on the resulting pathway graph, we evaluated the pathway identification procedure across multiple combinations of the importance threshold ($\tau$), the absolute pruning threshold ($\alpha$), and the relative pruning threshold ($\beta$).

\begin{figure}[htbp]
\centering
\begin{subfigure}{\textwidth}
\centering
\includegraphics[width=\linewidth]{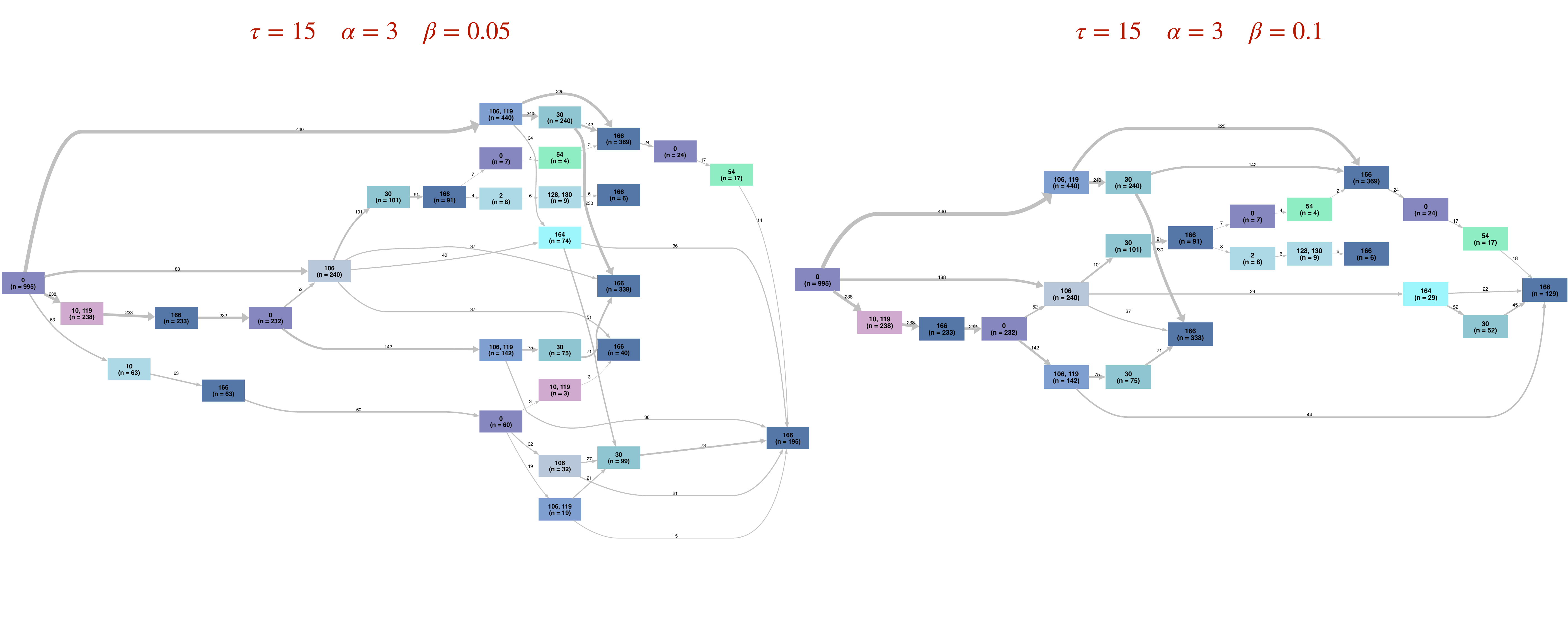}
\caption{}
\end{subfigure}

\vspace{0.5cm}

\begin{subfigure}{\textwidth}
\centering
\includegraphics[width=\linewidth]{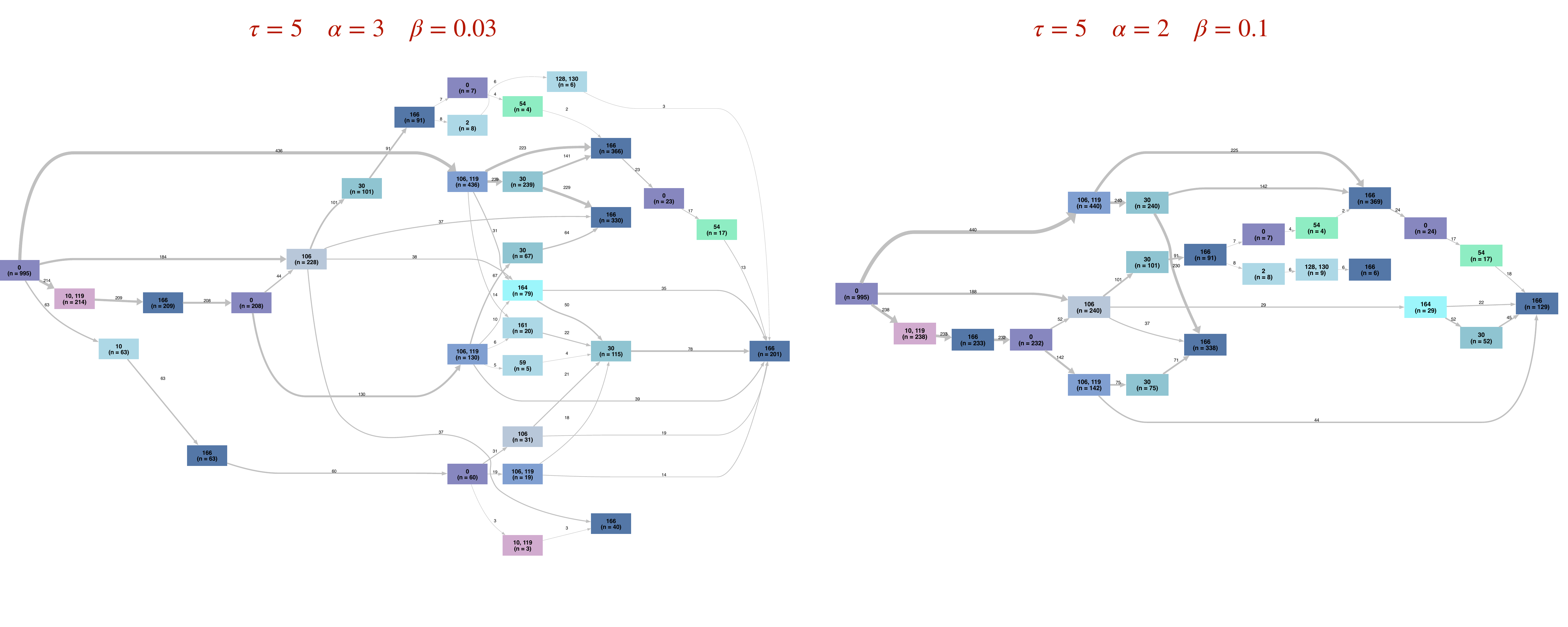}
\caption{}
\end{subfigure}
\caption{Sensitivity of the pathway graph with respect to different threshold parameters.}
\label{fig:supp_sensitivity}
\end{figure}

\begin{figure}[htbp]
\centering
\includegraphics[width=\textwidth]{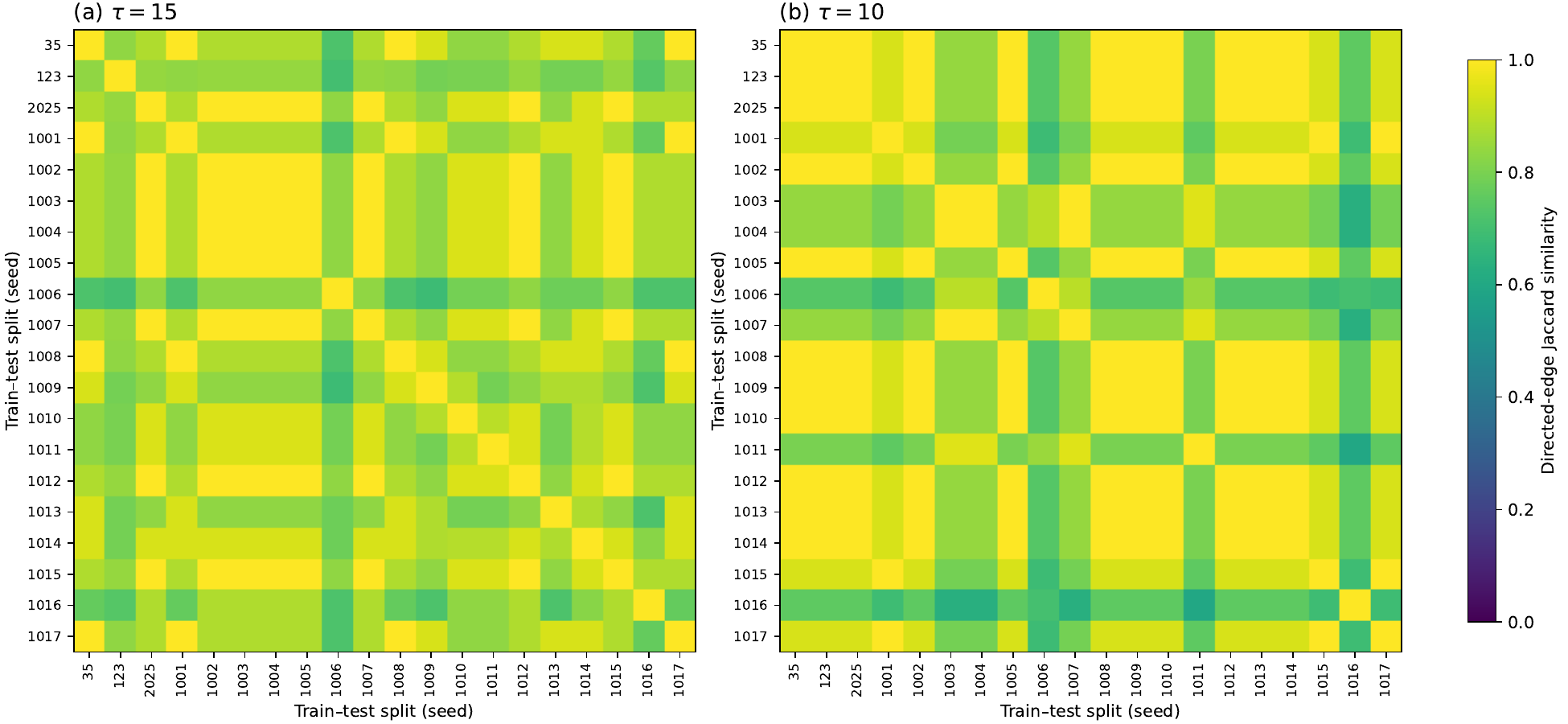}
\caption{Pairwise directed-edge Jaccard similarity of the displayed simplified pathway graphs across 20 train--test splits, for (a) $\tau = 15$ and (b) $\tau = 10$ (Section~\ref{sec:stability_training_splits}).}
\label{fig:graph_similarity}
\end{figure}

Figure~\ref{fig:supp_sensitivity} shows the simplified pathway graphs obtained for different parameter configurations. While lower thresholds retain additional peripheral nodes and higher thresholds yield more compact graphs, the dominant backbone structure of the pathway remains stable across parameter settings.

% Figure S4 is defined here (after Figure S3) so that the figure numbers used in the main text
% (Figure S2 = example graphs, Figure S3 = threshold sensitivity) are preserved.

\FloatBarrier

% =====================================================================================
% Reviewer 1, Comment 1 and Reviewer 2, Comment 3
\section{Number of pathway components and number of clusters}\label{sec:supp_components_clusters}
% =====================================================================================

\subsection{Pathway components}\label{sec:supp_components}
In admixture models for population genetics, the number of ancestral populations $K$ is estimated together with the components and must therefore be selected, for example by comparing likelihoods across values of $K$ \cite{pritchard2000}, by the rate of change of the likelihood ($\Delta K$; \cite{evanno2005}), or by cross-validation \cite{alexander2009}. In our framework, the components are not estimated freely but are given by the branches of the simplified pathway graph, so that $K$ follows from Step~1 rather than being selected separately. The question is therefore whether $K$ depends on the settings of Step~1.

The pathway components correspond to the branches of the simplified graph that start at the root (Section~3.2.1 of the main text). To assess whether their number depends on the analysis settings, we repeated Step~1 on the training patients of 20 train--test splits for five importance thresholds ($\tau = 5, 10, 15, 20, 30$), both for complete trajectories and for trajectories up to the first discharge. In all 200 settings, the simplified graph had exactly three branches, starting with $(106, 119)$, $(106)$, and $(10, 119)$.

For $\tau = 10$, the simplified graph contained a median of 9 root-to-terminal paths (range 8--12), all belonging to one of the three branches. A median of 15\% of training patients (109--136 of 796) had a first care step that was removed by pruning; they did not contribute to the estimation of the components but received admixture weights. Because nodes from later care stages can be merged into a branch during simplification, some patients were associated with two branches (in 10 of 20 splits; median 23 patients).

\subsection{Number of clusters}\label{sec:supp_clusters}

We clustered the admixture weights with $K_c = 2,\ldots,6$ clusters ($k$-means, 20 initializations, seed 42, as in Step~3) on the same 20 splits. For each $K_c$ we computed:
\begin{itemize}
    \item the internal indices of Table~1 of the main text, on the test patients;
    \item the cross-split stability of the partitions: the adjusted Rand index between the clusterings of all patients obtained from different training sets (190 split pairs);
    \item the share of patients in the smallest cluster;
    \item descriptively, Cram\'er's $V$ between the test-patient clusters and a subsequent recorded admission, with the analysis repeated on trajectories truncated at the first discharge (Section~4.3 of the main text).
\end{itemize}
The results are shown in Table~\ref{tab:clusters}.

\begin{table}[htbp]
\centering
\caption{Number of clusters $K_c$ in admixture space (SLSQP, $\tau = 10$); median over 20 train--test splits (stability: over 190 split pairs).}
\label{tab:clusters}
\small
\setlength{\tabcolsep}{4pt}
\renewcommand{\arraystretch}{1.15}
\begin{tabularx}{\textwidth}{
    c
    >{\centering\arraybackslash}X
    >{\centering\arraybackslash}X
    >{\centering\arraybackslash}X
    >{\centering\arraybackslash}X
    >{\centering\arraybackslash}X
    >{\centering\arraybackslash}X
}
\toprule
$K_c$ &
Silhouette &
\makecell{Calinski--\\Harabasz} &
\makecell{Davies--\\Bouldin} &
\makecell{Stability\\(ARI)} &
\makecell{Smallest\\cluster} &
\makecell{Cram\'er's\\$V$} \\
\midrule
2          & 0.58 & 227  & 0.66          & 0.67 & 43\% & 0.63 \\
\textbf{3} & 0.75 & 736  & \textbf{0.37} & 0.67 & 24\% & 0.84 \\
4          & 0.76 & 934  & 0.52          & 0.68 & 11\% & 0.84 \\
5          & 0.80 & 1228 & 0.43          & 0.71 & 7\%  & 0.85 \\
6          & 0.83 & 1795 & 0.43          & 0.73 & 4\%  & 0.85 \\
\bottomrule
\end{tabularx}
\end{table}

$K_c = 3$ gave the lowest Davies--Bouldin index. This also held for $\tau = 15$, for EM estimation and for trajectories up to the first discharge (eight settings in total). The silhouette and Calinski--Harabasz indices increased sharply from $K_c = 2$ to 3 and more slowly thereafter, as larger $K_c$ mainly split off small groups of patients with nearly identical admixture weights. Stability was similar for all $K_c$ (at $\tau = 15$ it was highest at $K_c = 3$: 0.81 vs.\ 0.74--0.76), and the association with subsequent admission did not increase beyond $K_c = 3$.

\FloatBarrier

% =====================================================================================
% Reviewer 1, Comment 4 and Reviewer 2, Comment 2
\section{Comparison with alternative approaches}\label{sec:supp_comparison}
% =====================================================================================

\subsection{Comparison with process mining}\label{sec:supp_process_mining}

\begin{figure}[htbp]
\centering
\begin{subfigure}[b]{\textwidth}
    \centering
    \includegraphics[width=0.9\textwidth]{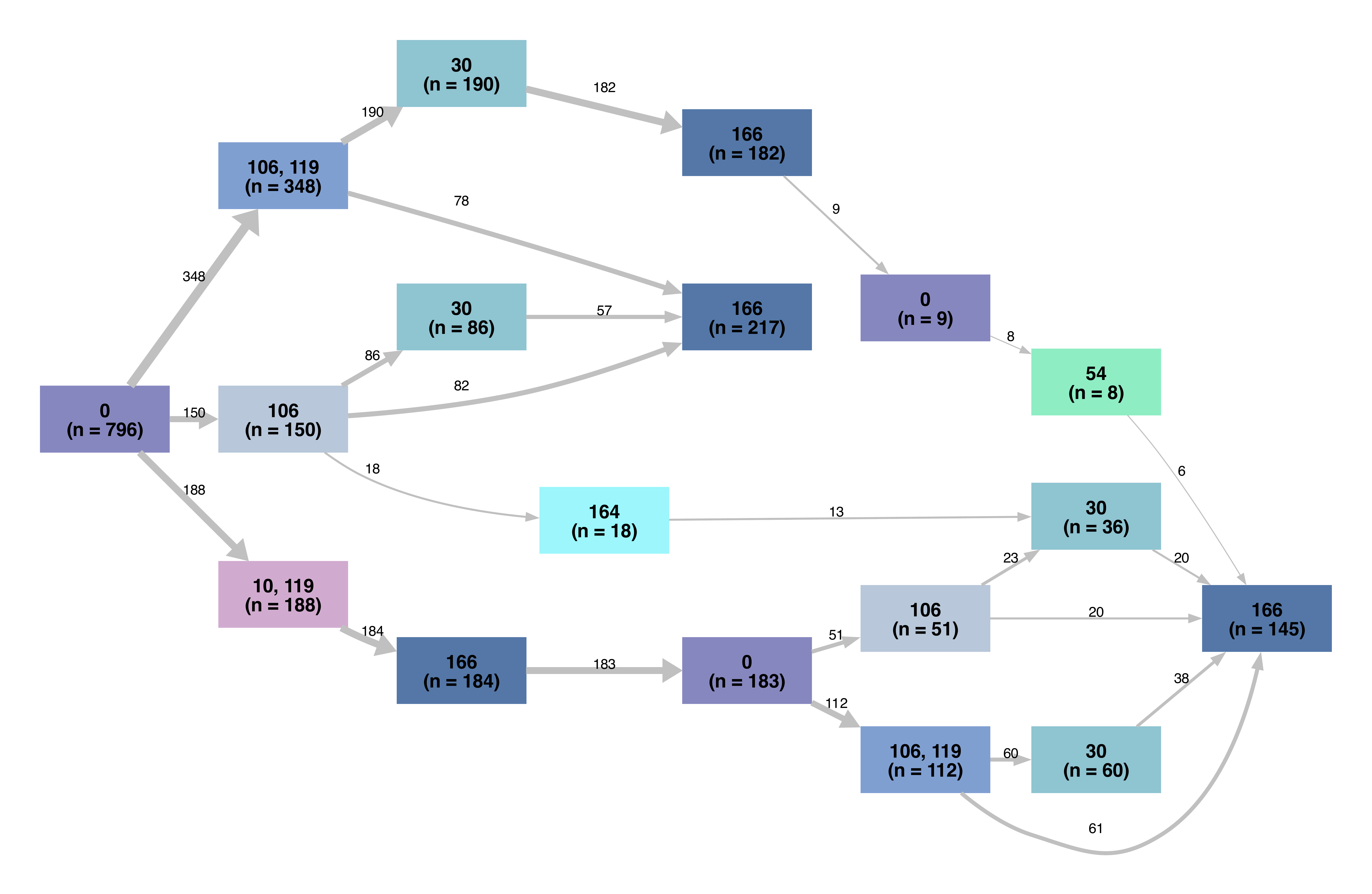}
    \caption{Typical-pathway graph (16 transition types)}
\end{subfigure}

\vspace{0.5cm}
\begin{subfigure}[b]{\textwidth}
    \centering
    \includegraphics[width=0.5\textwidth]{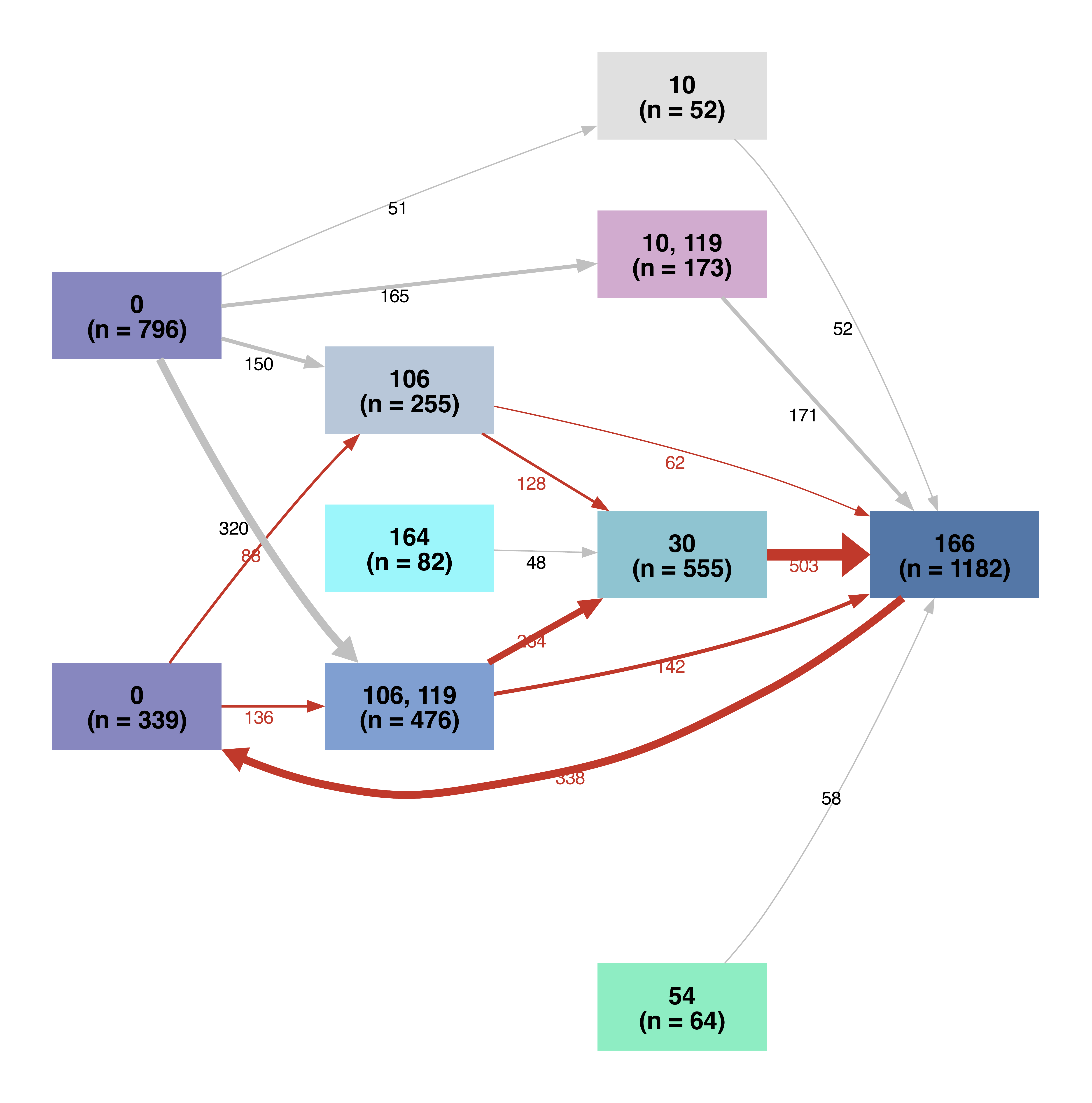}
    \caption{Directly-follows graph of the same size (16 transition types)}
\end{subfigure}
\caption{Typical-pathway graph (a) and directly-follows graph of the same size (b) for one split ($\tau = 10$). Red edges in (b) lie on a cycle, through which index stays and readmissions merge.}
\label{fig:pathway_vs_dfg}
\end{figure}

\begin{figure}[htbp]
\centering
\begin{subfigure}[b]{0.40\textwidth}
    \centering
    \includegraphics[width=\textwidth]{Figures/raw_dfg_threshold48_seed35.pdf}
    \caption{16 transition types (minimum count 48)}
\end{subfigure}
\hfill
\begin{subfigure}[b]{0.56\textwidth}
    \centering
    \includegraphics[width=\textwidth]{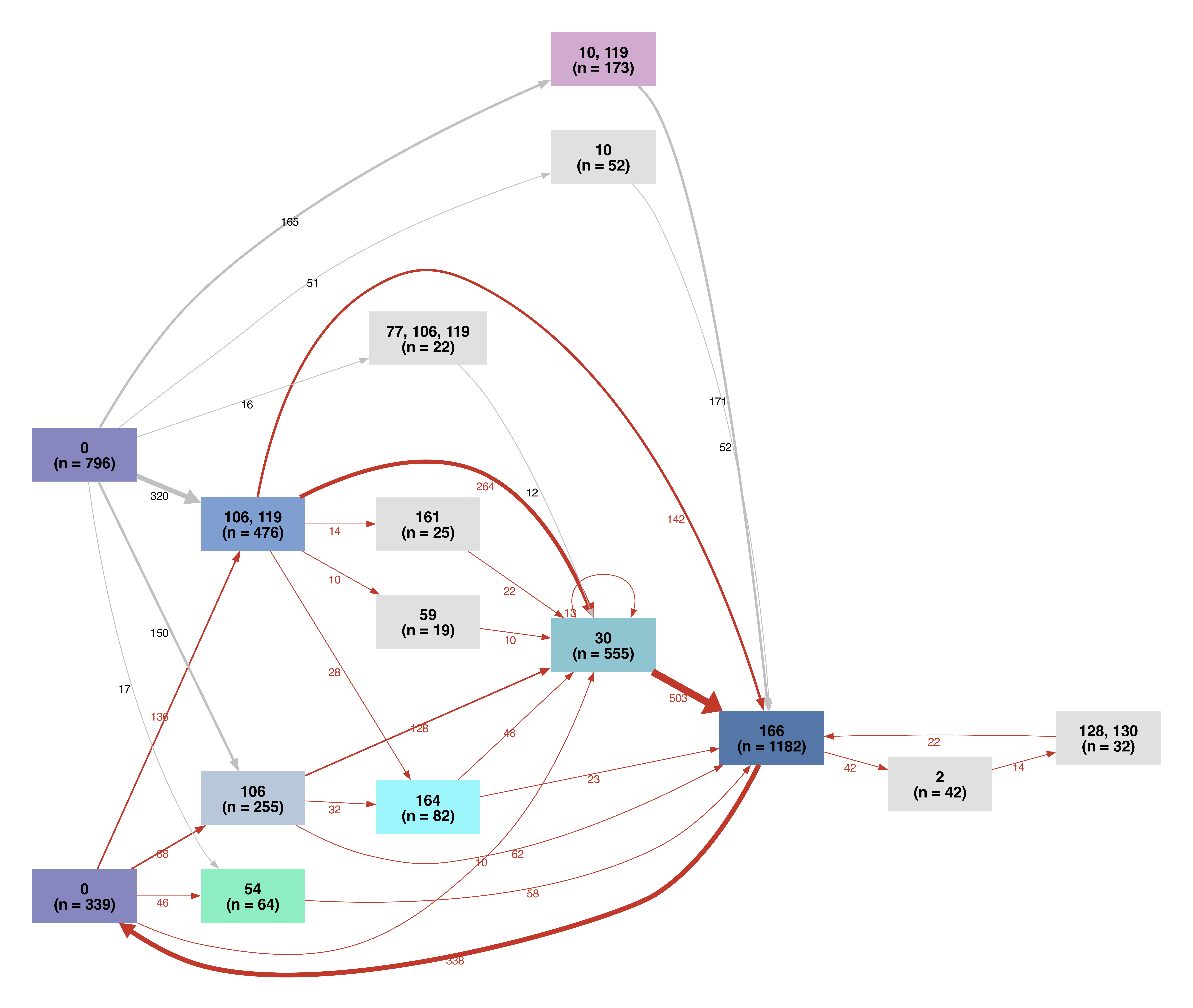}
    \caption{32 transition types (minimum count 10)}
\end{subfigure}

\vspace{0.4cm}
\begin{subfigure}[b]{\textwidth}
    \centering
    \includegraphics[width=\textwidth,height=0.45\textheight,keepaspectratio]{Figures/raw_dfg_threshold6_seed35.pdf}
    \caption{44 transition types (minimum count 6)}
\end{subfigure}
\caption{Directly-follows graphs for one split at decreasing frequency thresholds (16, 32 and 44 transition types). Red edges lie on a cycle.}
\label{fig:dfg_series}
\end{figure}

Figure~\ref{fig:pathway_vs_dfg} shows the typical-pathway graph and a directly-follows graph (DFG) of the same size for one split. The DFG threshold was the minimum transition frequency giving the number of transition types closest to that of the typical-pathway graph. With $\tau=15$, the DFG covered slightly more test transitions (72.9\% vs.\ 69.9\%) and was slightly more stable (0.94 vs.\ 0.88); it again contained cycles in 20/20 splits and unreachable nodes in 17/20. Less-filtered DFGs are shown in Figure~\ref{fig:dfg_series}. The Inductive Miner yielded models with low held-out precision (median token-replay precision 0.06--0.16).

\FloatBarrier

\subsection{Patient-level comparison}\label{sec:supp_patient_level}

All methods received the state sequences used in Step~2 (states with at least ten occurrences in the training set). Probabilistic methods used ten random restarts. For mixtures of hidden Markov models, the number of hidden states per component was selected from 2 to 5 by the Bayesian information criterion on the training patients; it selected 5 in all splits for complete trajectories and 4 for trajectories up to the first discharge. Clusters were fitted on training patients and test patients were assigned to the nearest cluster. Stability was computed on complete trajectories; for clinical association, all methods were refitted on trajectories truncated at the first discharge. The results for our framework in the main text refer to $\tau = 15$; with $\tau = 10$, its stability was 0.67 (ARI), with otherwise similar results.

\FloatBarrier

\end{document}